\documentclass[11pt]{article}
\usepackage{amsmath,amssymb}
\usepackage{graphicx,color}
\usepackage[hypertex]{hyperref}
\usepackage{srcltx}

\def\({\left(}
\def\){\right)}

\newcommand{\de}{\partial}
\newcommand{\be}{\begin{equation}}
\newcommand{\ba}{\begin{eqnarray}}
\newcommand{\ea}{\end{eqnarray}}
\newcommand{\ee}{\end{equation}}

\newcommand{\f}{\frac}
\newcommand{\s}{\sqrt}
\newcommand{\vp}{\varphi}

\newcommand{\ti}{\tilde}
\newcommand{\ap}{\alpha}

\newcommand{\ddd}{\cdot\cdot\cdot}
\newcommand{\no}{\nonumber \\}
\newcommand{\la}{\langle}
\newcommand{\lb}{\rangle}
\newcommand{\ep}{\epsilon}

 \def\de{\partial}

 \def\f {\frac}
 \def\ti{\tilde}
 \def\ap{\alpha}

 \def\ddd{\cdot\cdot\cdot}
 \def\no{\nonumber \\}

 \def\la{\langle}
 \def\lb{\rangle}
 \def\ep{\epsilon}

\begin{document}

\begin{titlepage}
\thispagestyle{empty}

\begin{flushright}
YITP-12-72
\\
IPMU12-0159
\\
\end{flushright}

\begin{center}
\noindent{\Large \textbf{
      Holographic Geometry of
      Entanglement Renormalization \\
      in Quantum Field Theories
}}\\
\vspace{2cm}

Masahiro Nozaki $^{a}$\footnote{e-mail:mnozaki@yukawa.kyoto-u.ac.jp},
Shinsei Ryu $^{b}$\footnote{e-mail:ryuu@illinois.edu} and
Tadashi Takayanagi $^{a,c}$\footnote{e-mail:takayana@yukawa.kyoto-u.ac.jp}
\vspace{1cm}

{\it
 $^{a}$Yukawa Institute for Theoretical Physics,
Kyoto University, \\
Kitashirakawa Oiwakecho, Sakyo-ku, Kyoto 606-8502, Japan\\
\vspace{0.2cm}
 $^{b}$Department of Physics, University of Illinois
 at Urbana-Champaign, \\
 1110 West Green St, Urbana IL 61801, USA \\
\vspace{0.2cm}
 $^{c}$Kavli Institute for the Physics and Mathematics of the Universe,\\
University of Tokyo, Kashiwa, Chiba 277-8582, Japan\\
 }

\vskip 2em
\end{center}

\begin{abstract}
We study a conjectured connection between AdS/CFT
and a real-space quantum renormalization group
scheme, the multi-scale entanglement
renormalization ansatz (MERA). By making a close contact with the holographic formula
of the entanglement entropy,
we propose a general definition of the metric in the MERA
in the extra holographic direction. The metric is formulated purely in terms of quantum field theoretical data. Using the continuum version of the MERA (cMERA),
we calculate this emergent holographic metric explicitly
for free scalar boson and free fermions theories,
and check that the metric so computed has the properties
expected from AdS/CFT.
We also discuss the cMERA in a time-dependent background induced
by quantum quench and estimate its corresponding metric.
\end{abstract}

\end{titlepage}

\newpage

\tableofcontents

\section{Introduction}

The holographic principle
\cite{Hol}
is a significant idea that
relates gravitational theories to those without gravity such as
quantum field theories.
The AdS/CFT correspondence is the most
well-understood example of the holography
\cite{Maldacena,GKP}.
On the one hand,
it allows us to study strongly coupled quantum systems by using
general relativity.
At the same,
the holographic dictionary, in principle,
allows us to
analyze quantum aspects of
gravity, which have been remaining as one of major challenges
in theoretical physics.
Our understanding on
the basic principle and mechanism of the holography
is, however, still limited.
To address
fundamental issues in quantum gravity,
such as
the creation of the universe or cosmology,
it is highly desirable to
develop a more general framework for the holography,
which allows us to go beyond
the anti de-Sitter spacetime
and explore non-trivial generalizations of AdS/CFT.

Recently, an interesting connection between AdS/CFT
and a real-space renormalization group
scheme in quantum many-body systems,
called the multi-scale entanglement
renormalization ansatz (MERA)
\cite{MERA,MERAR}, has been pointed out by Swingle
\cite{Swingle}.
Refer also to \cite{Ra,SWM,MS,BMR,Mat,ILN,MIH,Ok,MaK} for related recent works.

The MERA is originally proposed as an efficient
(numerical) method to find a
ground state for a given quantum many-body system.
A very brute-brash illustration of
the basic construction of the MERA may be given as follows.
(See the main text for more details.)
Suppose we want to find a ground state for an
interacting spin system on a one-dimensional lattice with $2^m$ sites.
To deal with the exponentially large Hilbert space,
we develop an iterative procedure;
starting from the true ground state with $2^m$ spins,
at each step of iteration,
we look for a description in terms of fewer
effective, degrees of freedom, by
coarse-graining the original spins.
In order for this ``naive'' application of the real
space renormalization group
idea to work for quantum systems, however,
quantum correlation between different spins need to be taken care of.
In the MERA the coarse-grain procedure is accompanied with
an unitary transformation, called ``disentangler'',
which is designed to remove quantum
entanglement at a given scale.
In the MERA, a quantum ground state is thus described by
a hierarchical structure that consists of
coarse-graining and disentanglers acting at different scales.
Ref.\ \cite{Swingle} proposed to
relate the above iterative steps of the MERA
to the extra dimension of the AdS.

(Alternatively, one can run the above procedure ``backward''
to develop an ``IR-to-UV'' picture;
In this picture, we start from a uncorrelated IR state
to describe the our target (ground) state at UV
by using the coarse graining transformations iteratively.
Along these steps, we use the ``disentangler'' to generate entanglements
between spins, in such a way that the UV state correctly
mimics the true ground state as closely as possible.
See the main text for more details.)

An important hint for the connection between AdS/CFT
and the MERA is obtained by comparing how
the entanglement entropy is estimated in both cases
\cite{Area}.
(For reviews of the entanglement entropy
in quantum many-body systems, see, e.g.,
\cite{Ereview,CCreview,CHreview,Lreview}).
In the classical gravity limit of AdS/CFT,
we can calculate the entanglement entropy $S_A$ for a subregion
$A$ from the minimal area surface \cite{RT}
(for reviews see \cite{HEEreview}):
\begin{align}
S_A=\f{\mbox{Area}(\gamma_A)}{4G_N},
\label{HEE}
\end{align}
where $G_N$ is the Newton constant of the AdS
gravity; $\gamma_A$ is the minimal area surface which ends on the
boundary of $A$
at the AdS boundary.
The holographic formula (\ref{HEE}) has passed many tests:
the analytical agreements in AdS$_3$/CFT$_2$ \cite{RT};
the strong subadditivity
\cite{SSA};
the phase transition when the subsystem $A$ is disconnected \cite{Hed};
the agreements
of the coefficient of logarithmic terms \cite{RT,Log,CHM};
moreover, the proof of (\ref{HEE}) has been given in \cite{CHM} when
the subregion $A$
is a round sphere.
Strictly speaking, this
prescription can be applied only to static systems.
However, in general
time-dependent backgrounds we can employ the covariant
formulation in \cite{HRT}.

The holographic formula (\ref{HEE}) already suggests a possibility
of reconstructing the bulk spacetime in gravity from the quantum field
theoretic data \cite{HRT}.
If we know the ground state of a given
quantum field theory or quantum many body system,
then we can in
principle calculate the entanglement entropy $S_A$ for any choice of
subsystem $A$.
Since $S_A$ is related to the area of a minimal
surface in its gravity dual (such as the AdS),
it is natural to
expect that we can reconstruct the metric of this gravitational
spacetime by considering all possible choices of the subsystem $A$.
Refer to \cite{Ham,Hub} for several developments in this direction.

The structure of the entanglement included in
the (candidate) ground state in the MERA
is set by its coarse-graining structure
and the disentanglers.
We can thus estimate the entanglement entropy $S_A$
essentially
by counting how many disentanglers,
which may be operative at different iterative steps,
are relevant for a given region $A$.
Such counting gives us a bound for the entanglement entropy,
which looks quite similar to the holographic formula (\ref{HEE})
as pointed out in \cite{Swingle}.
(See the main text for more details.)

Though traditionally the extra dimension of
a AdS has often
been identified with the energy scale of the Wilsonian renormalization
group flow in momentum space \cite{BVV,HP,FLR},
the calculation of the entanglement entropy suggests that AdS/CFT
fits nicely with the real space renormalization group scheme,
especially with the MERA.
This is simply
because the AdS metric is defined in the real space and it is not
straightforward to see the
structures in momentum space of its dual conformal field theories geometrically.

The purpose of this paper is to make the interpretation of AdS/CFT
as the entanglement renormalization more concrete by looking
closely at the holographic geometry of a given quantum field theory.
In order to match with
AdS/CFT we need to take the continuum limit of the MERA. This is
recently studied in \cite{cMERA} and is called the cMERA
(continuous MERA). We will use
this formalism and relate it to {AdS/CFT}.
In particular, we propose a proper
definition of metric in the extra dimension purely in terms of data
of quantum field theories for translational invariant states. This
will be a first step to understand an emergence of gravitational
spacetime from quantum field theories based on the idea of
holography. For earlier works on emergent gravity via holography
refer to, e.g., \cite{Lee,Ra}.

This paper is organized as follows: In section two, we introduce the idea of
the MERA and cMERA.
In section three, we formulate
the cMERA for a free scalar field theory so that it includes a series of excited states.
In section four, we study the connection
between AdS/CFT and the (c)MERA;
we introduce a holographic metric in the extra direction.
In section five, we give analogous results for a free fermion theory in two dimension.
In section six, we summarize our results discuss future problems.

\section{Entanglement Renormalization: MERA and cMERA}

In this section, we introduce and review the MERA
(the multi-scale entanglement renormalization ansatz)
\cite{MERA},
and its continuum counter part, the cMERA
(the continuous multi-scale entanglement renormalization ansatz)
\cite{cMERA}.
They were proposed as an efficient (numerical) method to find a ground state
of quantum many-body systems by the variational principle,
and can be considered as an implementation of
the real space renormalization group.
Since the MERA can be thought of
as an example of more general framework,
called the tensor network algorithms or tensor network methods,
let us start by describing tensor network algorithms in general, and in particular,
one of their simplest examples;
the matrix product states (MPS) and
the density matrix renormalization group (DMRG).

\subsection{Tensor Network Algorithms}

A major challenge in quantum many-body problems
is their exponentially-large Hilbert space.
A successful (numerical) algorithm for quantum many-body systems
should provide an efficient way to explore/access the physically relevant
subspace of the total Hilbert space.
(For example,
Quantum Monte Carlo attempts to achieve this by importance sampling.)
Different phases in the quantum many-body systems
may be classified/categorized in terms of their computational complexity,
or, equivalently,
the scaling of the entanglement entropy as a function of the size
of a spatial region of interest.
The strategy of Tensor Network Algorithms
(Tensor Network Methods)
to attack quantum many-body problems
is to (i) target
a given class of computational complexity
and to (ii) find, within the given computational class,
an optimal quantum state which
most faithfully describes the true ground state.
In this way, Tensor Network Methods avoid
the exponential blow-up of the Hilbert space dimension as a function of
the system size.

In the tensor network methods,
a quantum state $|\Psi\lb$ is
described in terms of a set of tensors.
Each tensor can be of arbitrary rank and can carry multiple indices.
Some of the indices represent physical degrees of freedom
(such as quantum spin degrees of freedom
defined at each site of a give lattice system),
while the others represent unphysical, auxiliary degrees of freedom
that are introduced for the purpose of efficiently writing down a ground state
-- see below for more details.
A physical wavefunction (quantum state) is represented
as a product of tensors
with properly contracted indices in the auxiliary spaces.

\begin{figure}[ttt]
   \begin{center}
     \includegraphics[height=4cm]{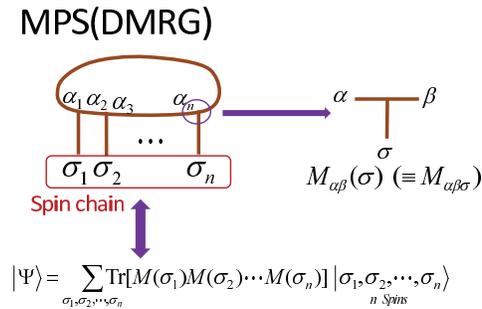}
   \end{center}
   \caption{The matrix product state (MPS) for a quantum spin chain.
The upper left diagram represents a tensor network for the MPS,
which consists of ``tripods'' shown in the upper left diagram.
The corresponding quantum state is shown below.
\label{fig:MPSS} }
\end{figure}

\begin{figure}[ttt]
   \begin{center}
     \includegraphics[height=4cm]{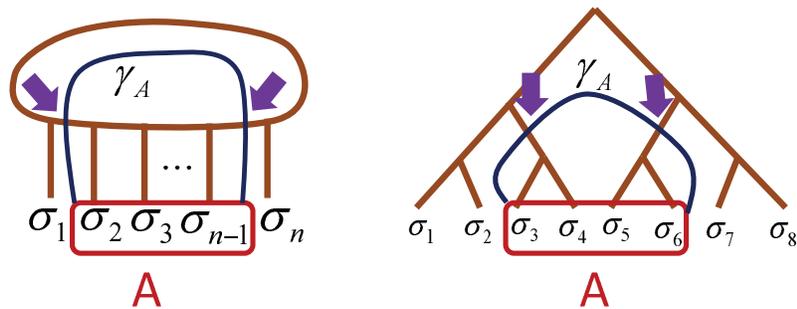}
   \end{center}
   \caption{
     The tensor network diagrams and the estimation
     of the entanglement entropy $S_A$ for the matrix product state (MPS)
     (left)
     and for the tree tensor network (TTN) (right).
     \label{fig:TTNE}
    }
\end{figure}

Tensor network representations of a quantum state can roughly be described
in terms of their ``structure'' and ``variational parameters''.
The former includes the rank of tensors involved and how these tensors
are contracted.
This can often be described in terms of a network diagram,
as in Figs.\ \ref{fig:MPSS}, \ref{fig:TTNE}, \ref{fig:MERAE}.
This structural aspect
largely sets the computational class of possible quantum
states that the tensor network can describe.
On the other hand, the latter includes
the dimensions of auxiliary spaces and actual
numerical values of tensor elements. They can be optimized,
within a given structure of the tensor network,
to find a best tensor network state which best describes the target quantum state.

The matrix product state (MPS) \cite{MPS} or the density matrix renormalization group (DMRG) \cite{White} is a canonical example of Tensor Network Algorithms in one spatial dimension,
and one of the most successful algorithms for
many-body quantum systems. Consider, as an example, a quantum spin
chain with $n$ $S=1/2$ degrees of freedom.
A quantum state
 $|\Psi\lb$ is described by specifying the coefficients (wavefunction)
$c(\sigma_1,\sigma_2,\ldots,\sigma_n)$:
\begin{align}
|\Psi\lb=\sum_{\sigma_1,\sigma_2,\ldots,\sigma_n}
c(\sigma_1,\sigma_2,\ldots,\sigma_n)~|\sigma_1,\sigma_2,\ldots,\sigma_n\lb,
\end{align}
where $\sigma_{i}=0,1$ (spin up and down).
In the MPS in one spatial dimension,
we construct the coefficient $c(\sigma_1,\sigma_2,\ldots,\sigma_n)$ as
a product of tensors
(tensor network)
constructed from, as a building block,
a tensor $M_{\alpha\beta}(\sigma)$ with three indices.
Each tensor $M$ can pictorially be represented as a ``junction'' or ``tripod''
as in the upper right part of Fig.\ \ref{fig:MPSS}.
The tensor $M_{\alpha\beta}(\sigma)$ has one physical or spin index,
and two auxiliary or ``internal''
indices $\ap,\beta$ that take values $1,2,\ldots,J$;
$J$ is one of (variational) parameters of the tensor network.
The coefficients
$c(\sigma_1,\sigma_2,\ldots,\sigma_n)$ in the MPS
are given
by considering $n$ such ``tripods'',
which are, in simplest cases, all
identical,
and contracting indices for the auxiliary spaces
for a given set of physical indices.
Thus,
$c(\sigma_1,\sigma_2,\ldots,\sigma_n)$ in this construction
is given by a product of $J\times J$
matrices $M_{\ap\beta}(\sigma)$ as
\begin{align}
c(\sigma_1,\sigma_2,\ldots,\sigma_n)=\mbox{Tr}[M(\sigma_1)M(\sigma_2)\ddd M(\sigma_n)].
\end{align}
This is pictorially represented as in the upper left part of Fig.\ \ref{fig:MPSS};
When two legs of junctions ($=M$) are connected
in the diagram, we contract the indices between them.
In DMRG,
  a candidate  for the ground state of the system is represented
  as a MPS, parameterized by a set of
  matrices $M_{\ap\beta}(\sigma)$,
  and then it is variationally optimized.
  The candidate state can be systematically improved
  by increasing the dimension $J$ of the auxiliary space.

By introducing higher rank tensors and considering tensors connected (contracted) via
more complex network, the MPS can be extended to higher dimensions.
Such an generalized framework is called projected entangled-pair states (PEPS)
\cite{PEPS,NO}.

Let us now ask, focusing again on systems in one spatial dimension,
what kinds of quantum states can be represented as a MPS;
what is the computational class of wavefunctions represented by the MPS?
To answer this question, let us estimate
the entanglement entropy $S_A$ for
MPS wavefunctions when bipartitioning the system
into two subsystems $A$ and $B$ and taking a partial trace over $B$.
One finds it is bounded as
\begin{align}
S_A\leq 2\log J.
\label{boundd}
\end{align}
This can be understood as follows.
Upon bipartitioning, the only parts of the tensor network (MPS)
that contribute to the entanglement entropy are
the ``bonds'' which are ``cut'' by bipartitioning.
These bonds are not a coupling in the physical Hamiltonian,
but one of the legs of a tripod $M$ which is connecting two
auxiliary indices.
The maximal contribution to the entanglement entropy from each bond is given by $\log J$.
In this way we obtain the bound (\ref{boundd}).
The factor of two comes from the two ends of the region $A$.
For later purposes,
we introduce a curve $\gamma_A$
to specify this partitioning;
it has its ends on the interface of the subregions $A$ and $B$,
and it intersects with the two bonds in the tensor network;
See the left panel in Fig.\ \ref{fig:TTNE}.

The entanglement entropy $S_A$ can thus be made arbitrary large if we increase
the dimension $J$ of the auxiliary space.
However, it does not scale with the size of the subregion $A$.
On the other hand, it is well-known that the entanglement entropy in a one dimensional critical
system (relativistic CFT)
is proportional to $\log L$, where $L$ is the size of the subsystem $A$
\cite{HLW,Cardy}.
Therefore the bound (\ref{boundd}) shows that MPS ansatz cannot realize
the large amount of entanglement required to simulate
quantum critical systems.
(On the other hand,
it is known that, the ground state of any one-dimensional gapped system
can faithfully represented by the MPS if $J$ is sufficiently large.)

\subsection{MERA}
\label{subsec MERA}

\begin{figure}[ttt]
   \begin{center}
     \includegraphics[height=4cm]{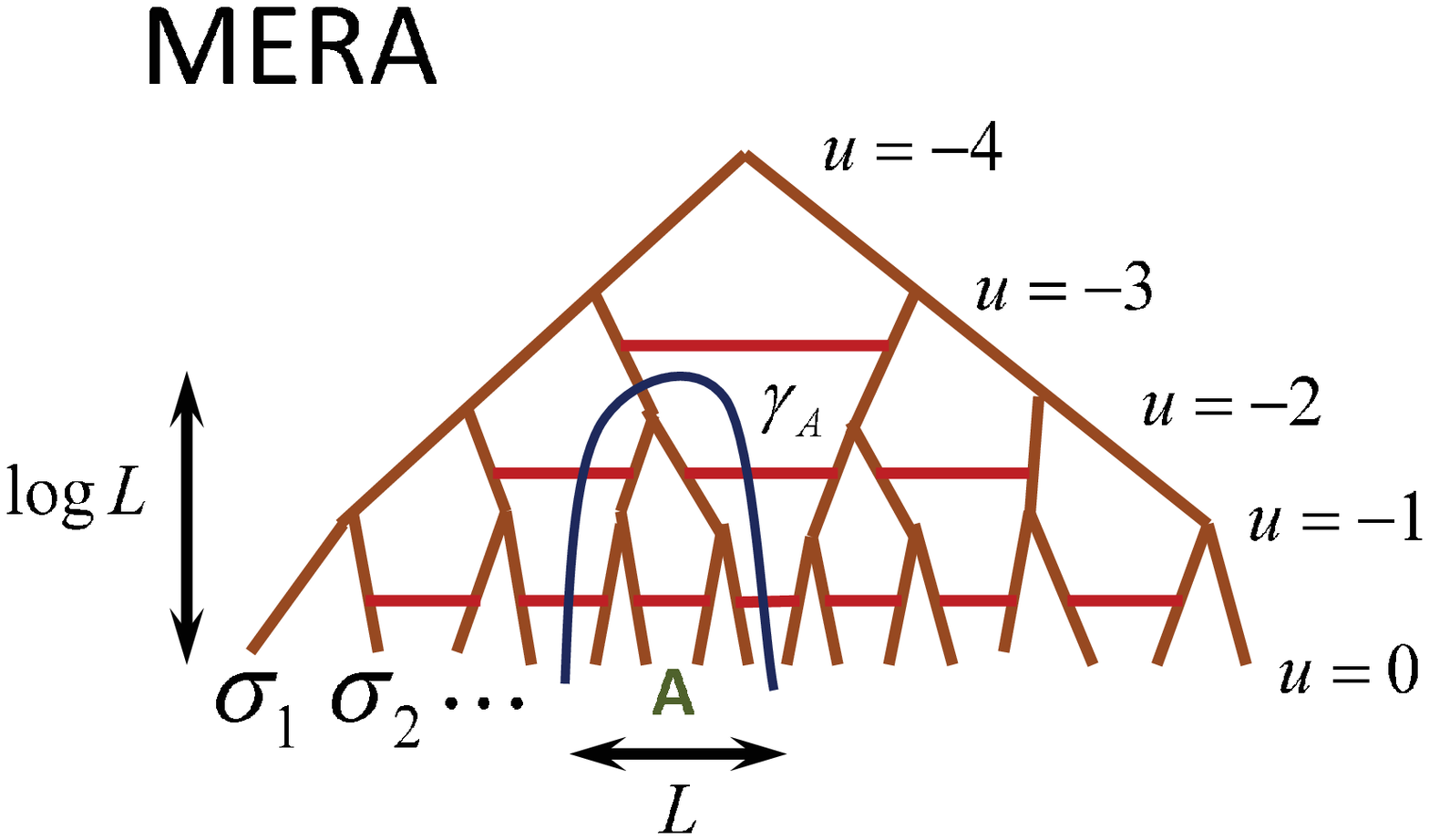}
\hspace{1cm}
     \includegraphics[height=3cm]{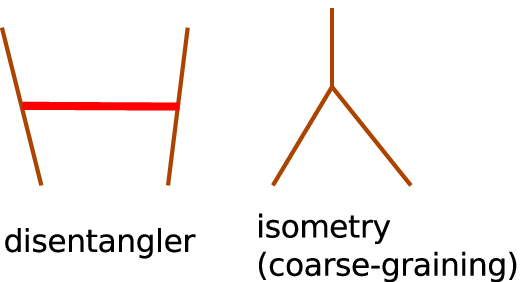}
   \end{center}
   \caption{
     (Left) The tensor network for the MERA and the estimation of
     the entanglement entropy.
     The brown trees describes the coarse-graining
     of the original spin chain.
     The red horizontal bonds describe the disentanglers,
     which is an unitary transformation acting on each pair of two spins.
     It is clear from this picture that we can estimate the entanglement
     entropy as
     $\mbox{Min}_{\gamma_A}\left[{\#}\mbox{Bonds}(\gamma_A)\right]\sim \log L$.
     (Right) The MERA tensor network consists of disentanglers (``tetrapods'')
     and isometries implementing coarse-graining (``tripod'').
     \label{fig:MERAE}}
\end{figure}

In order to look for an alternative to the MPS,
which would work better for quantum critical systems,
we need to modify the structure of the tensor network.
The tensor network for the MERA
is
different from the MPS
and
has a layered structure
as pictorially represented
in Fig.\ \ref{fig:MERAE}.
We label layers by an integer $u$ according to their depth or generation.
We take $u=0,-1,-2,...,-\infty$ for later convenience.
The original spin chain is located on the $0$-th layer ($u=0$),
which is at the bottom of the tensor network in Fig.\ \ref{fig:MERAE}.
The second layer ($u=-1$) is obtained by combining two spins into
a single one by a linear map (isometry),
which is regarded as a coarse graining procedure or
equally the scale transformation;
the layered structure is motivated by the real space
renormalization group idea.
We can repeatedly apply this procedure as many as one wishes.
If the original system has $n$ spins,
there are $n\cdot 2^{u}$ spins on the layer $u$.

The coarse graining is,
however, not the only ingredient of the MERA.
In fact,
if it were so, the tensor network would look like
the right panel of Fig.\ \ref{fig:TTNE};
This tensor network construction is called the tree tensor network (TTN).
The entanglement entropy of the TTN is,
again, bounded by a formula like (\ref{boundd}),
and does not scale as the function of the size of the subregion $A$.
This can easily be understood pictorially
from the right panel in Fig.\ \ref{fig:TTNE}.
To remedy this situation, we are lead to add extra bonds
called ``disentanglers'',
as explained in Fig.\ \ref{fig:MERAE}.
Physically,
this means that we perform an appropriate unitary transformation
on each of the Hilbert spaces of two spins.
To summarize,
the coarse graining procedure together with
the disentangler
defines the tensor network of the MERA
\cite{MERA}.
The tensor network consists of
tripods representing the coarse graining transformation
(called ``isometries'')
and
tetrapods which represent ``disentanglers''.
As before,
in order to obtain the ground state (numerically),
we  minimize the total energy by optimizing
the parameters in isometries and disentanglers.
(See Fig.\ \ref{fig:MERAE}.)

  In order to better understand the effects of disentanglers,
let us now estimate the entanglement entropy in the MERA for a subsystem $A$ with $L$ spins.
 For a general tensor network state, the entanglement entropy $S_A$
is bounded as
\be
S_A\leq {\#}\mbox{Bonds}(\gamma_A) \cdot \log J,
\ee
where ${\#}\mbox{Bonds}(\gamma_A)$ is
the number of bonds intersected by $\gamma_A$ in the tensor network.
  Notice that in general there are various choices for $\gamma_A$
  and ${\#}\mbox{Bonds}(\gamma_A)$ depends on the choice.
  Therefore we need to minimize this to obtain the most strict bound which can be saturated by the tensor network as
\be
S_A\leq \mbox{Min}_{\gamma_A}\left[{\#}\mbox{Bonds}(\gamma_A)\right] \cdot \log J.
\label{bondm}
\ee
In the MERA for a spin chain,
it is easy to estimate the curve $\gamma_A$ which leads to the minimum entropy and
we obtain the following bound as explained in Fig.\ \ref{fig:MERAE}:
\be
 S_A\leq c \cdot \log L,  \label{boundm}
 \ee
 where $c$ is a numerical constant of order one. In this way, the entanglement structure
 of the MERA is consistent with that of quantum critical systems.

It is also worth while mentioning
that it is straightforward to find higher dimensional versions of
the MERA for e.g. spin systems in two and three spacial dimensions by changing the structure of
the coarse-graining and the disentanglers accordingly.

\subsection{cMERA}

For applications to quantum field theories, it is desirable to consider a continuum formulation of the MERA. One such formulation is recently explored in \cite{cMERA} and is called
the cMERA (continuous MERA). This formulation is also useful in order to make clearer the connection between AdS/CFT and entanglement renormalization as we will explain in a later part of this paper.

Assume that a QFT with a Hamiltonian is given. We need to impose a UV cut off
$\Lambda=\f{1}{\ep}$, where $\ep$ can be identified as the lattice constant.
We denote the Hilbert space defined by the fields with the UV cut off $\Lambda$
as ${\cal H}_\Lambda$.
As in the MERA \cite{MERA,MERAR}, we will perform coarse graining procedures.
We thus consider a one-parameter family of wave functions (or states)
\be
|\Psi(u)\lb\in {\cal H}_\Lambda,
\ee
where $u$ represents the length scale of interest.
We take $u$
such that the momentum $k$ is effectively cut off as  $|k|\leq \Lambda e^{u}$.
We can regard $u$ as the continuum analogue of $u$ introduced in the MERA to specify the
layers. It is important that we always work with the same microscopic Hilbert space ${\cal H}_\Lambda$.
The UV limit and the IR limit are defined to be
$u=u_{\mathrm{UV}}=0$ and
$u=u_{\mathrm{IR}}\to -\infty$.

We define the states in the IR and UV limit as follows:
\begin{align}
|\Psi(u_{\mathrm{IR}})\lb &\equiv |\Omega\lb,
\\
|\Psi(u_{\mathrm{UV}})\lb &\equiv |\Psi\lb.
\end{align}
The UV state $|\Psi\lb$ is what describes the system we are studying, typically the ground state for a given Hamiltonian. In this paper we always assume that the system is described by a pure state. On the other hand, the IR state $|\Omega\lb$ is a trivial
state in that there is no quantum entanglement between any spacial regions in that system and the entanglement entropy vanishes $S_A=0$ for any subsystem $A$ defined in a real space.
We can relate $|\Psi(u)\lb$ and $|\Omega\lb$ by a unitary transformation:
\begin{align}
  |\Psi(u)\lb=U(u,u_{\mathrm{IR}})|\Omega\lb.
  \label{uvc}
\end{align}
Equivalently we can represent $|\Psi(u)\lb$ in terms of the UV state:
\begin{align}
  |\Psi\lb=U(0,u)|\Psi(u)\lb.
  \label{uzc}
\end{align}

Now we express the unitary transformation as
\be
U(u_1,u_2)=
P
\exp \left[
  -i\int^{u_1}_{u_2}(K(u)+L)du
\right],
\label{uuc}
\ee
where $K(u)$ and $L$ are the continuum analogue of the disentangler and the scale transformation
(coarse graining)
in the cMERA, respectively \cite{cMERA}.
The symbol $P$ means a path-ordering which puts
all operators with smaller $u$
to the right. For later convenience we also define $\ti{P}$ as the one with the opposite order.

We require that the IR state $|\Omega\lb$ is invariant under the scale transformation $L$:
\be
L|\Omega\lb=0.  \label{linv}
\ee
This is because the IR state does not have any quantum entanglement and each spacial point behaves independently.
On the other hand, following the discrete MERA,
the operator $K(u)$ is designed to generate
the entanglement for the modes with wave vectors $|k|\leq \Lambda e^u$,
as becomes clearer from the later arguments.
Therefore, (\ref{uvc}) shows that the UV state $|\Psi\lb$ is constructed from
the non-entangled state $|\Omega\lb$ by repeating the addition of the quantum entanglement and the scale transformation as $u$ varies from $-\infty$ to $0$.
If we view this in an opposite way from the UV to IR limit,
at each step
we disentangle the system by $K(u)$
and then do a coarse-graining by $L$.

At first sight, the construction of quantum states by the cMERA looks quite different from the MERA.
In particular, in the MERA the dimension of the Hilbert space (the number of spins)
is reduced by half in each step of coarse graining, while in the cMERA the dimension of
the Hilbert space is preserved.
It is, however, possible to formulate the MERA in terms of a unitary evolution
as in the cMERA.
This can be done by simply adding a fixed dummy spin state $|0\lb$
in each step of coarse graining in the MERA
so that it keeps of the size of Hilbert space.
The coarse-graining procedure in the cMERA can be more properly understood as the continuum
limit of this latter definition of the MERA.

By moving to the 'Heisenberg picture', we can define an operator $O$ at scale $u$ as
\begin{align}
O(u)=U(0,u)^{-1}\cdot O\cdot U(0,u),
\end{align}
where $O$ is the operator in the UV limit. This satisfies
\be
\la\Psi|O|\Psi\lb=\la \Psi(u)|O(u)|\Psi(u)\lb.
\ee

Finally it is useful to rewrite $U$
in the `interaction picture':
\be
U(u_1,u_2)=e^{-iu_1 L}P e^{-i\int^{u_1}_{u_2}\hat{K}(u)du}\cdot e^{iu_2L},
\ee
where we defined
\be
\hat{K}(u)=e^{iuL}\cdot K(u)\cdot e^{-iuL}. \label{defkh}
\ee
In this way we find for the UV state:
\begin{align}
  |\Psi\lb=P e^{-i\int^{0}_{u_{\mathrm{IR}}}\hat{K}(u)du}|\Omega\lb.
  \label{xxx}
\end{align}

In general, the operator $K(u)$ is chosen so that it creates the quantum entanglement
for the scale below the UV cut off, i.e., $|k| \leq \Lambda $.
The effective operator $\hat{K}(u)$ when we factor out the scale transformation is defined by
(\ref{defkh}) and thus $\hat{K}(u)$ generates the entanglement for the scale
$|k|\leq \Lambda e^u$ in the expression (\ref{xxx}) as advertised.

For a finite $u$, we get:
\be
|\Psi(u)\lb=e^{-iuL}\cdot P e^{-i\int^{u}_{u_{\mathrm{IR}}}\hat{K}(s)ds}|\Omega\lb. \label{xy}
\ee

For later purpose, it is also useful to define
\be
|\Phi(u)\lb =e^{iuL}|\Psi(u)\lb=P e^{-i\int^{u}_{u_{\mathrm{IR}}}\hat{K}(s)ds}|\Omega\lb. \label{vy}
\ee
The physical meaning of the state $|\Phi(u)\lb$ is as follows: as $u$ varies from $-\infty$ to $u$, we add the entanglement by the operator
$K(u)$. For the scale higher than $u$, we just perform the scale transformation $L$ until it reaches the UV limit. This defines the state $|\Phi(u)\lb$.

\section{Analysis of cMERA in Free Scalar Field Theory}

  Here we analyze the cMERA in a free scalar field theory, which is the main example
in this paper. We will closely follow the formulation in \cite{cMERA} (the appendix of version 1)
and generalize this construction so that it can describe a class of excited states with a generic dispersion relation.

\subsection{Formulation of cMERA in Free Scalar Field Theory}

Consider the $(d+1)$-dimensional free scalar field with the general dispersion relation
$\ep_k$. In momentum space, the Hamiltonian is given by
\be
H=\f{1}{2}\int d^dk\,  [\pi(k)\pi(-k)+\ep_k^2\cdot \phi(k)\phi(-k)]. \label{haml}
\ee
The results for a standard massive relativistic scalar field is obtained by setting
$\ep_k=\s{k^2+m^2}$. We can express $\phi(k)$ and $\pi(k)$ in terms of creation and annihilation operators
\be
\phi(k)=\f{a_k+a^{\dagger}_{-k}}{\s{2\ep_k}}, \ \ \ \ \pi(k)=\s{2\ep_k}\left(\f{a_k-a^{\dagger}_{-k}}{2i}\right).
\ee
The commutation relation between $a_k$ and $a^{\dagger}_k$ is given by
\be
[a_k,a^{\dagger}_p]=\delta^d(k-p),
\ee
which is equivalent to
\be
[\phi(k),\pi(p)]=i\delta^d(k+p).
\ee

First, we define the IR state $|\Omega\lb$ by
\begin{align}
  \left(\s{M}\phi(x)+\f{i}{\s{M}}\pi(x)\right)|\Omega\lb=0,
  \label{IRS}
\end{align}
where $M$ is a constant, which is taken to be order of the UV cut off $\Lambda$ as we will confirm later.
Note that for this state $|\Omega\lb$, the entanglement entropy $S_A$ is indeed vanishing for any subsystem $A$ because
all modes for any $x$ are decoupled from each other. It satisfies
\begin{align}
  \la\Omega|\phi(k)\phi(k')|\Omega\lb =\f{1}{2M}\delta^{d}(k+k'),
  \quad
 \la\Omega|\pi(k)\pi(k')|\Omega\lb=\f{M}{2}\delta^d(k+k').
\end{align}
In the oscillator expression, $|\Omega\lb$ is defined by the property
\begin{align}
  (\ap_k a_k+\beta_k a^{\dagger}_{-k})|\Omega\lb=0,
  \label{unents}
\end{align}
where
\begin{align}
  \ap_k=\f{1}{2}\left(\s{\f{M}{\ep_k}}+\s{\f{\ep_k}{M}}\right),
  \quad
  \beta_k=\f{1}{2}\left(\s{\f{M}{\ep_k}}-\s{\f{\ep_k}{M}}\right).
\end{align}

The IR state $|\Omega\lb$ is invariant
under the (non-relativistic) scale transformation $L$:
\ba
&& e^{-iuL}\phi(x)e^{iuL}= e^{\f{d}{2}u}\phi(e^ux),\ \ \ \  e^{-iuL}\phi(k)e^{iuL}= e^{-\f{d}{2}u}\phi(e^{-u}k),\no
&& e^{-iuL}\pi(x)e^{iuL}= e^{\f{d}{2}u}\pi(e^ux),\ \ \ \ e^{-iuL}\pi(k)e^{iuL}= e^{-\f{d}{2}u}\pi(e^{-u}k).
\ea

Note that $L$ differs from the standard (relativistic) scale transformation $L'$ which is given by
\ba
&& e^{-iuL'}\phi(x)e^{iuL'}= e^{\f{d-1}{2}u}\phi(e^ux),\ \ \ \  e^{-iuL'}\phi(k)e^{iuL'}= e^{-\f{d+1}{2}u}\phi(e^{-u}k),\no
&& e^{-iuL'}\pi(x)e^{iuL'}= e^{\f{d+1}{2}u}\pi(e^ux),\ \ \ \ e^{-iuL'}\pi(k)e^{iuL'}= e^{-\f{d-1}{2}u}\pi(e^{-u}k).
\ea

Now, we introduce the disentangler as follows \cite{cMERA}
\be
K(u)=\f{1}{2}\int d^dk \left[
g(k,u)(\phi(k)\pi(-k)
+
\pi(k)\phi(-k))\right].
\label{dise}
\ee
We assume\footnote{More generally we can consider a disentangler which is not
s-wave, where $g(k,u)$
depends not only $|k|$ but also on the vector $k$ as $g(k,u)=a_{i_1i_2\ddd i_l}(|k|,u)\cdot k^{i_1}k^{i_2}\ddd k^{i_l}$. The free fermion theory corresponds to $l=1$ example in this sense as
we will seen in section 5. In the light of holography, such a generalization will correspond to the excitations of higher spin fields in the dual higher spin gravity theory as we will comment in the final section.} that the $k$ dependence of the real valued function $g(k,u)$ is s-wave i.e. only depends on $|k|$.
The function $g(k,u)$ is assumed to have the following form
\begin{align}
g(k,u)=\chi(u)\cdot \Gamma \left(|k|/\Lambda\right),
\end{align}
where $\Gamma(x)=\theta(1-|x|)$ is the cut off function
($\theta(x)$ is the step function); $\chi(u) $ is a real valued function.
Although at this point (\ref{dise}) is an ansatz to find a
ground state, we will show later that it produces the exact ground state.
Then the transformation $U(0,u)$ defined by (\ref{uuc}) acts as
\begin{align}
 U(0,u)^{-1}\phi(k)U(0,u)&= e^{-f(k,u)}e^{-\f{u}{2}d}\phi(e^{-u}k),
  \nonumber \\
  U(0,u)^{-1}\pi(k)U(0,u) &= e^{f(k,u)}e^{-\f{u}{2}d}\pi(e^{-u}k).
\end{align}
Notice that in the interaction picture (\ref{xxx}) we find
\begin{align}
  \hat{K}(u)
  &=\f{1}{2}\int d^d k\, e^{du} \left[g(k,u)\phi(ke^u)\pi(-ke^u)+g(k,u)\pi(ke^u)\phi(-ke^u)\right]
  \nonumber \\
&=\f{1}{2}\int d^d k \left[g(ke^{-u},u)\phi(k)\pi(-k)+g(ke^{-u},u)\pi(k)\phi(-k)\right]. \label{disein}
\end{align}
The function $f(k,u)$ satisfies
\be
\f{\de f(k,u)}{\de u}=g(ke^{-u},u),
\ee
which is solved as
\be
f(k,u)=\int^u_0 ds\  g(ke^{-s},s).
\ee
Note that the final expression in (\ref{disein}) shows that the momentum integral in $\hat{K}(u)$ has the cut off $|k|\leq \Lambda e^{s}$.

Finally we apply the variational principle and minimize the energy \cite{cMERA}.
The total energy $E$ is given by
\begin{align}
  E&=\la \Psi|H|\Psi\lb =\la \Omega|H(u_{\mathrm{IR}})|\Omega\lb
  \nonumber \\
  &= \la \Omega| \int d^d k\,  \f{1}{2}
  \Big[e^{2f(k,u_{\mathrm{IR}})}e^{-u_{\mathrm{IR}}d}\pi(ke^{-u_{\mathrm{IR}}})\pi(-ke^{-u_{\mathrm{IR}}})
  \nonumber \\
  &\qquad \qquad +\ep_k^2\cdot
  e^{-2f(k,u_{\mathrm{IR}})}e^{-u_{\mathrm{IR}}d}\phi(ke^{-u_{\mathrm{IR}}})\phi(-ke^{-u_{\mathrm{IR}}}) \Big]|\Omega \lb
  \nonumber \\
  &= \int d^dx \int d^d k\, \f{1}{4}\left[e^{2f(k,u_{\mathrm{IR}})}M+\f{\ep_k^2}{M}e^{-2f(k,u_{\mathrm{IR}})}\right].
\end{align}
We require the variation of $E$ with respect to $\chi(u)$ for any $u$ is vanishing:
\be
\f{\delta E}{\delta \chi(u)}=\int d^d x\int d^d k \left( e^{2f(k,u_{\mathrm{IR}})}M-\f{\ep_k^2}{M}e^{-2f(k,u_{\mathrm{IR}})}\right)
\Gamma(|k|e^{-u}/\Lambda)=0.
\ee
Thus we find
\be
f(k,u_{\mathrm{IR}})=\f{1}{2}\log \f{\ep_k}{M}, \ \ \
(|k|<\Lambda). \label{condf}
\ee
By using
\be
f(k,u_{\mathrm{IR}})=\int^{u_{\mathrm{IR}}}_0 g(ke^{-s},s)ds=\int^{-\log \Lambda/|k|}_0 \chi(s)ds,
\ee
we obtain
\be \chi(u)=\f{1}{2}\cdot \left(\f{|k|\de_{|k|}
\ep_k}{\ep_k}\right)\Biggr|_{|k|=\Lambda e^u}. \label{gdp} \ee
This characterizes the ground state $|\Phi\lb$ of the free scalar field theory given by the Hamiltonian (\ref{haml}).

In particular, for the free scalar field theory with a mass $m$ \cite{cMERA} we obtain
\be
\chi(u)=\f{1}{2}\cdot \f{e^{2u}}{e^{2u}+m^2/\Lambda^2}, \ \ \ M=\s{\Lambda^2+m^2}. \label{chs}
\ee
The function $f(k,u)$ is given by
\begin{align}
  f(k,u)
  =
  \left\{
    \begin{array}{ll}
      \displaystyle
    \f{1}{4}\log \f{m^2+e^{2u}\Lambda^2}{m^2+\Lambda^2}, & (|k|<\Lambda e^{u})
    \nonumber \\
    \nonumber \\
    \displaystyle
    \f{1}{4}\log \f{k^2+m^2}{m^2+\Lambda^2}. & (|k|>\Lambda e^{u})
    \end{array}
  \right.
\end{align}

Some comments are in order here.
In the CFT (massless limit),
we always have $\chi(u)=\f{1}{2}$.
Thus one can show that the total operation $K+L$ at each scale
coincides with the relativistic scale transformation (dilatation) $L'$ of the CFT.
In the massive case, we have $\chi (u)\simeq \f{1}{2}$ in the UV region, i.e.,
$e^{u}\gg \f{m}{\Lambda}$. On the other hand, in the IR region $e^{u} \ll \f{m}{\Lambda}$, we have $\chi(u)\simeq 0$ and the unitary transformation acts trivially, corresponding to the absence of mass gap.

It is also intriguing to consider non-standard scalar fields. If we assume the dispersion relation $\ep_k\propto k^{\nu}$, which corresponds to a Lifshitz theory (anisotropic scale invariant theory) with the dynamical exponent $\nu$, we again find that $\chi(u)$ takes a constant value
\be
\chi(u)=\f{\nu}{2}. \label{mlif}
\ee

\subsection{Excited States in cMERA}

Here we would like to describe a class of excited states in the cMERA given by the coherent
states:
\be
(A_k a_k+B_k a^{\dagger}_{-k})|\Psi_{ex}\lb=0, \label{conda}
\ee
where we normalize
\be
|A_k|^2-|B_k|^2=1. \label{norab}
\ee
Notice that the definition of $(A_k,B_k)$ still has the ambiguity of phase factor.
The ground state corresponds to the choice $B_k=0$.


We want to relate the UV state $|\Psi_{ex}\lb$ to
the unentangled IR state $|\Omega\lb$ via the
unitary transformation
\be
|\Psi_{ex}\lb=Pe^{-i\int^0_{u_{\mathrm{IR}}}\hat{K}_{\psi}(u)du}|\Omega\lb, \label{req}
\ee
as in (\ref{xxx}).
In terms of oscillators, we assume the form
\be
\hat{K}_\psi(u)=\f{i}{2}\int d^dk\, \left(g_k(u) a^{\dagger}_k a^{\dagger}_{-k}-g^*_k(u) a_k a_{-k}\right),
\ee
where the function $g_k$ is taken to be the form
\be
g_k(u)=g(u)\Gamma(|k|e^{-u}/\Lambda).
\ee
For example, $\hat{K}(u)$ in (\ref{disein})
can be written in this form with the identification $
g_k(u)=\Gamma(|k|e^{-u}/\Lambda)\chi(u)$. In general $g_k(u)$ takes complex values and $\hat{K}_\psi(u)$ cannot be cast into the form (\ref{disein}).

To proceed, it is useful to look at how the unitary transformation in (\ref{req}) acts on the creation and annihilation operators.
It is given by a linear transformation of the
following form:
\begin{align}
  P e^{-i\int^u_{u_{\mathrm{IR}}}\hat{K}_\psi(u)du}
  \cdot
\left(\begin{array}{c}
a_k  \\
a^{\dagger}_{-k}
\end{array}\right)
\cdot
\ti{P} e^{i\int^u_{u_{\mathrm{IR}}}\hat{K}_\psi(u)du}
=
\left(\begin{array}{cc}
  p_k(u) & q_k(u) \\
  q^*_k(u) & p^*_k(u)
\end{array}\right)
\cdot
\left(\begin{array}{c}
a_k \\
a^{\dagger}_{-k}
\end{array}\right).
\label{trsm}
\end{align}
Here $p_k(u)$ and $q_k(u)$ are given in terms of $g_k(u)$ in the following way:
\begin{align}
M_k(u)
\equiv
\left(\begin{array}{cc}
  p_k(u) & q_k(u) \\
  q^*_k(u) & p^*_k(u)
\end{array}\right)
=
\ti{P}
\exp\left(-\int^u_{u_{\mathrm{IR}}}du\, G_k(u)\right),  \label{msde}
\end{align}
where the matrix $G_k(u)$ is defined by \ba G_k(u)=
\left(\begin{array}{cc}
  0 & g_k(u) \\
  g^*_k(u) & 0
\end{array}\right)
=\Gamma(|k|e^{-u}/\Lambda)\left(\begin{array}{cc}
  0 & g(u) \\
  g^*(u) & 0
\end{array}\right).
\label{gks} \ea The matrix $M_k(u)$ defined by (\ref{msde})
satisfies \be \f{dM_k(u)}{du}=-M_k(u)G_k(u). \label{uuu} \ee $M_k(u)$ clearly
satisfies $|p_k(u)|^2-|q_k(u)|^2=1$ and preserves
the commutation relations between $a_k$ and $a^{\dagger}_k$.

Eventually, for the UV state $|\Psi_{ex}\lb$, we find that
$(A_k,B_k)$ in the definition (\ref{conda}) is related to the IR
values $(\ap_k,\beta_k)$ by \be (A_k,B_k)=(\ap_{k},\beta_{k})\cdot
M_{k}(0).  \label{abtr} \ee

As is clear from (\ref{gks}) in general we can decompose $G_k(s)$
into two pieces
\be
G_k(u)=\ti{G}(u)\Gamma(|k|e^{-u}/\Lambda).
\label{pic}
\ee
This allows us to rewrite $M_k(0)$
\ba M_k(0) =
\ti{P} \cdot \exp\left(-\int^0_{\log\f{|k|}{\Lambda}}du\,
\ti{G}(u)\right).
\label{mmat}
\ea
Notice the relation
\be
\ti{G}_{11}(u)=\ti{G}^*_{22}(u)=-\ti{G}^*_{11}(u),\ \ \ \ \
\ti{G}_{12}(u)=\ti{G}^*_{21}(u). \ee

It is also useful to extend the previous construction to the state at scale $u$
\be
|\Psi_{ex}(u)\lb=e^{-iuL}\cdot Pe^{-i\int^u_{u_{\mathrm{IR}}}\hat{K}_{\psi}(s)ds}|\Omega\lb=
e^{-iuL}|\Phi_{ex}(u)\lb, \label{reuq}
\ee
as we did in (\ref{xy}). Since $|\Psi_{ex}(u)\lb$ is obtained from the scale transformation of $|\Phi_{ex}(u)\lb$, we need only to consider the relation between $|\Phi_{ex}(u)\lb$ and $|\Omega\lb$. This can be found from (\ref{trsm}) as follows:
\ba (A_k(u),B_k(u))
=(\ap_{k},\beta_{k})\cdot M_{k}(u), \label{xc} \ea
where we assumed that $|\Phi_{ex}(u)\lb$  satisfies $(A_k(u) a_k+B_k(u) a^{\dagger}_{-k})|\Phi_{ex}(u)\lb=0$.

Before we go on we have a few remarks. Notice that $M_k(u)$ satisfies
\be
M_k(u)^\dagger \sigma_3 M_k(u)=\sigma_3,  \ \ \  \det M_k(u)=1, \label{pp}
\ee
which says that $M_k(u)$ belongs to $SU(1,1)$.

The parametrization (\ref{gks}) covers only two out of three generators of $SU(1,1)$. The remaining one is found to be the form
\be
e^{i\delta_k(u)\sigma_3} \in SU(1,1). \label{rrr}
\ee
If we fix $(\ap_k,\beta_k)$ and ask if we can find $M_k(u)$ for an arbitrarily given pair
$(A_k,B_k)$, $M_k(u)$ should have three parameters. However, if we remember that $(A_k,B_k)$
is defined up to a phase factor, $M_k(u)$ only needs to have two parameters. Therefore, the ansatz (\ref{gks}) is enough and we can ignore the one (\ref{rrr}).

\subsubsection{Ground State}

The ground state of the cMERA corresponds to $A_k=1$ and $B_k=0$
and thus we find from (\ref{xc})
\ba
M_k(0)=
\left(\begin{array}{cc}
  \ap_k & -\beta_k \\
  -\beta_k & \ap_k
\end{array}\right).
\ea
If we assume that $g_k(u)$ takes only real values i.e. $g_k=(g_k)^*$,
then we find from (\ref{gks})
\begin{align}
M_k(u)=
\left(\begin{array}{cc}
  \cosh\int^u_{u_{\mathrm{IR}}}du\, g_k(u) & -\sinh\int^u_{u_{\mathrm{IR}}}du\, g_k(u) \\
-\sinh\int^u_{u_{\mathrm{IR}}}du\, g_k(u)   & \cosh\int^u_{u_{\mathrm{IR}}}du\, g_k(u)
\end{array}\right). \label{grmk}
\end{align}

By comparing between these in the UV limit $u=0$, we need to require
\be
e^{\int^0_{u_{\mathrm{IR}}}du g_k(u)}=\s{\f{M}{\ep_k}}.
\ee
This is equivalent to (\ref{condf}) and therefore indeed it agrees with the result in the
previous section. The explicit form of $g_k$ is given by
\be
g_k(u)=\chi(u)\Gamma(|k|e^{-u}/\Lambda),
\ee
where $\chi(u)$ is defined in (\ref{gdp}). In this example of the ground state, the matrix
$G_k(s)$ (\ref{gks}) explicitly reads
\ba
G_k(s)=
\chi(u)\Gamma(|k|e^{-u}/\Lambda)
 \left(\begin{array}{cc}
   0 & 1 \\
   1 & 0
 \end{array}\right).
\label{gksz}
\ea

\subsection{Time-dependent Excited State}

We can find $p_k(0)$ and $q_k(0)$ (or equivalently $M_k(0)$)
from $A_k$
and $B_k$ as follows: \ba && p_k(0)=\ap_kA_k-\beta_k B^*_k, \no  &&
q_k(0)=-\beta_k A^*_k+\ap_k B_k. \ea By taking the derivative with
respect to the momentum $k$, we obtain from (\ref{mmat})
\begin{align}
  \ti{G}\left(\log\f{|k|}{\Lambda}\right)
  &=|k|\f{dM(0)}{d|k|}\cdot M^{-1}(0)
\nonumber \\
&= \left(\begin{array}{cc}
  |k|(\de_{|k|} p)p^*-|k|(\de_{|k|} q) q^{*} & |k|p(\de_{|k|} q)-|k|q(\de_{|k|} p) \\
  -|k|(\de_{|k|} p^{*})q^*+|k|p^*(\de_{|k|} q^{*}) & -|k|q(\de_{|k|} q^{*})+|k|p(\de_{|k|} p^{*})
\end{array}\right),
\label{gmat}
\end{align}
where we omit, for simplicity, the arguments of $p_k(0)$ and $q_k(0)$ as $p$ and
$q$. After we calculate $\ti{G}(u)$ from (\ref{gmat}), we can
eventually compute $M_k(s)$ following (\ref{msde}) in principle.

We would like to analyze a class of excited states given by
(\ref{conda}). In particular, we assume that at time $t=0$,
the ratio $A_k/B_k$ is real. This includes the important examples
of quantum quenches discussed later. By taking into its time
evolution, $A_k$ and $B_k$ are given by the following form: \ba &&
A_k=\s{1+a_k^2}\cdot e^{i\ep_k t+i\theta_k}, \no && B_k=a_k\cdot
e^{-i\ep_k t+i\theta_k}, \label{abqq} \ea where $a_k$ is taken to be an
arbitrary real valued function of $k$. The phase $\theta_k$ is a
redundant parameter on which the UV state $|\Psi_{ex}\lb$ does not
depend. We choose $\theta_k$ such that $\ti{G}_{11}=0$.
Considering the
limit $M\sim \Lambda \gg \ep_k$, we obtain
\begin{align}
\de_{|k|}\theta_k&=t
  \left(2a_k\s{1+a_k^2}\cos2\theta_k-1-2a_k^2\right)\de_{|k|}\ep_k-\f{\de_{|k|} a_k\cdot\sin2\theta_k}{\s{1+a_k^2}}, \label{mqq} \\
\ti{G}_{12}&=g\left(\log\f{|k|}{\Lambda}\right)=\f{|k|\de_{|k|}\ep_k}{2\ep_k}
\left(1+4ta_k\ep_k\s{1+a_k^2}\sin2\theta_k\right)+\f{|k|\de_{|k|} a_k\cdot\cos2\theta_k}{\s{1+a_k^2}}.
\label{mqc}
\end{align}
Note that this shows that the function $g(u)$ in this final expression (\ref{mqc}) is real valued in the low energy region $-u\gg 1$. This property is no longer true in the high energy region $u\sim 0$, where similar calculations lead to a slightly complicated expression.
Below we focus on the former region as that is enough to find the essential property we want. Since $g(u)$ is real, we can find the expression of $M_k(s)$ and it is given by (\ref{grmk}) with $\chi(u)$ replaced with $g(u)$.

 One important property which we can find from (\ref{mqc}) is that at late time $g(u)$ increases linearly in general as long as $a_k\neq 0$. This agrees with the fact that the entanglement entropy generally grows linearly under a time evolution after a
 strong excitation (or called quantum quench)
 \cite{Quench}
 as we explain in the next section.

One may notice that the choice of $\theta_k$ has an ambiguity which corresponds to an integration constant when we solve (\ref{mqq}). In other words, $\ti{G}_{12}$ or equally $g(u)$ depend on the choice of $\theta_k$. This is because even though the UV state $|\Psi_{ex}\lb$ defined in  (\ref{conda}) is independent from the choice of $\theta_k$, the middle energy state $|\Psi_{ex}(u)\lb$ (or equally $|\Phi_{ex}(u)\lb$) does depend on $\theta_k$. There does not seem to exist any uniquely preferable choice of
$\theta_k$. This means that the definition of the cMERA for a given excited state is not
unique at least within the current approach. We will later discuss the holographic interpretation of this ambiguity and argue that it corresponds to the choice of time slices extended into the extra dimension.

\subsection{Global Quantum Quenches}

One convenient physical process to create an strongly excited state is the
global quantum quench where the system uniformly gets excited by a
sudden shift of some parameters such as mass or coupling constants \cite{Quench,CCreview}.
Here we consider the global quench triggered by the shift of mass of
the free scalar. Especially, we choose it such that the mass is
$m_0$ for $t<0$ and it suddenly goes to $m$ at $t=0$ and later.

By requiring that the field and its momentum are continuous at
$t=0$, we find the system for $t>0$ coincides with the excited state
(\ref{conda}) defined by (\ref{mqc}) with the specific choice of
$a_k$
\begin{align}
  a_k
  =\f{1}{2}
  \left[
    \left(\f{k^2+m_0^2}{k^2+m^2}\right)^\f{1}{4}
    -\left(\f{k^2+m^2}{k^2+m_0^2}\right)^\f{1}{4}
  \right].
\label{abq}
\end{align}

\begin{figure}[ttt]
   \begin{center}
     \includegraphics[height=5cm]{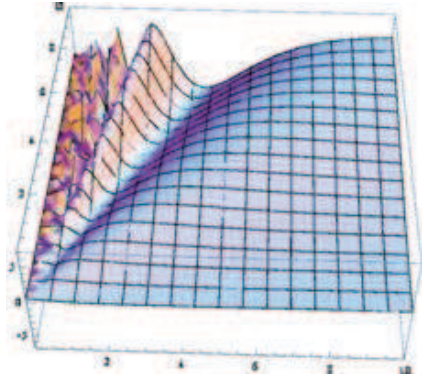}
   \end{center}
   \caption{
     The time-evolution of $\ti{G}_{12}=g(u)$ after a quantum quench, computed for $m=0.1$ and $m_0=1$ using the same approximation in (\ref{gfu}). We fixed the phase ambiguity as
     $\theta_0(t)=0$.
     The horizontal and vertical coordinate correspond to $1/|k|=\ep e^{-u}$ and $t$, respectively.
     The former, $1/|k|$, can be interpreted as the $z$ coordinate in the AdS as we explain
      in section 4.  With this interpretation,
    we can observe, qualitatively, that the excitations
    induced by the quantum quench are within the light-cone:
    $z<t$ in the AdS.
     \label{fig:QunechMERA}}
\end{figure}

Let us consider the case where $m=0$ i.e. $\ep_k=|k|$ so that we obtain a CFT after
the quench. The differential equations (\ref{mqq}) and (\ref{mqc}) get simpler in the region $m_0\ll |k| \ll \Lambda$ and we can solve as
\begin{align}
  & \theta_k\simeq -|k|t+\theta_0(t),
  \nonumber \\
& \ti{G}_{12}=
g\left(\log\f{|k|}{\Lambda}\right)\simeq
\f{1}{2}-\f{m^2_0}{2|k|}t\sin(2|k|t-\theta_0(t))-\f{m_0^2}{2k^2}\cos(2|k|t-\theta_0(t)), \label{gfu}
\end{align}
where $\theta_0(t)$ is an undetermined integration constant as discussed in the previous subsection. We plotted the form of $\ti{G}_{12}=g(u)$ as a function of $1/|k|$ and $t$ by choosing $\theta_0(t)=0$ in Fig.\ \ref{fig:QunechMERA}, where introduced the IR cut off
by turning on $m$ slightly so that the divergence of $a_k$ (\ref{abq}) in the IR limit $|k|\to 0$ is regularized.

\section{AdS/CFT and Entanglement Renormalization}

As we have explained, the MERA and the cMERA are a useful tool in quantum many-body problems and quantum field
theories, and can be viewed as an efficient implementation of the real space renormalization group idea.
As conjectured in \cite{Swingle}
(see also \cite{TNG,cMERA} for later related arguments),
it is natural to expect some relation between
the MERA/cMERA and the AdS/CFT correspondence,
where in the latter the extra direction in a AdS
is interpreted as the energy scale
of the renormalization group flow.
The purpose of this section is to materialize this interpretation
by introducing a proper metric structure for the tensor networks in the MERA/cMERA.

\subsection{MERA and AdS/CFT Correspondence}

\begin{figure}[ttt]
   \begin{center}
     \includegraphics[height=4cm]{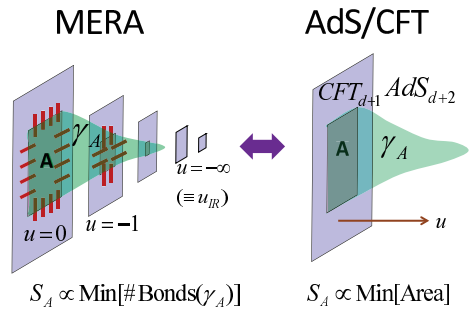}
   \end{center}
   \caption{
     The parallelism between
     the calculations of the entanglement entropy $S_A$
     in the MERA (left)
     and AdS/CFT (right). The green surface represents $\gamma_A$ in both pictures. The red bonds
     in the left denote the disentanglers.    \label{fig:AdSMERAE}
   }
\end{figure}

The holographic formula (\ref{HEE})
provides us with a way to compute the entanglement entropy in a QFT
from the bulk geometry.
Since {AdS/CFT} is a duality,
it would be possible to ``read backward''
the holographic dictionary,
and infer the bulk geometry from entanglement in the QFT.
This strategy could be extended to a generic class of QFTs,
and in particular to QFTs for which we can find a ground state
by the MERA.
Therefore,
let us start by remembering
how the entanglement entropy can be estimated for
a wavefunction represented in terms of a MERA network;
it was discussed in Sec.\ \ref{subsec MERA}.
The entanglement entropy in the MERA network is bounded as
(\ref{boundm}) (and (\ref{bondm}) for the case of one-dimensional
critical systems).
In particular,
if each bond is maximally entangled
the equality in (\ref{boundm}) should holds.
This bound shows that the entanglement entropy
can be estimated
by the minimum of the number of bonds cut
among choices of divisions $\gamma_A$
for a given tensor network
(see the left picture in Fig.\ \ref{fig:AdSMERAE}).
This estimation of the entanglement entropy in MERA is
very analogous to that of the holographic entanglement entropy
\cite{Swingle}.

This identification between the surface $\gamma_A$ in the MERA
and $\gamma_A$ in the holographic formula
suggests the relation
\be
{\#}\mbox{Bonds}
\simeq
\f{\mbox{Area}(\gamma_A)}{G_N},
\label{bondar}
\ee
assuming that the bonds are maximally entangled. Note that the minimization of both sides
agree with each other consistently.
Here $\simeq$ means the equality up to a numerical factor of order one.
This relation (\ref{bondar}) suggests, at least, a strong correlation
between the entanglement in a QFT
and the bulk geometry (if it exits).
The relation (\ref{bondar}) can also be thought of as
a natural generalization of the standard microscopic interpretation of black hole entropy,
in that it says that
the area of a minimal
surface (measured in the unit of Planck length) corresponds
to the number of entangling bonds.

The comparison between the holographic formula and the entanglement entropy
bound in the MERA further suggests
the extra direction in the AdS is identified
with the $u$ in the MERA which counts
the layers of each coarse-grained spin system
(Fig.\ \ref{fig:AdSMERAE}).
To be more precise,
let us write the $(d+2)$-dimensional AdS metric in the form
\be
ds_{\mathrm{AdS}}^2=du^2+\f{e^{2u}}{\ep^2}(d\vec{x}^2-dt^2)=\f{dz^2-dt^2+d\vec{x}^2}{z^2}, \label{adsmet}
\ee
where $\vec{x}$ describes the coordinate of $R^d$;
$\ep$ is the lattice spacing of the spin system.
We normalize the radius of the AdS to be unity for simplicity;
$z=\ep\cdot e^{-u}$ is the standard radial coordinate of Poincare AdS.
The extra direction $u$ in the AdS
is identified with $u$ in the MERA
when it is representing a ground state of a conformal invariant system
(they are equal up to a proportionality constant as we explain below).

In the above example of a conformal invariant system,
the component of the metric $g_{uu}$ is constant.
This is in line with the fact that in the scale invariant the MERA,
the structure of the tensor network, including the disentanglers, do not
change as we go deeper into the IR region (by increasing the layer index $u$).
For quantum field theories which are not conformally invariant,
the MERA network would evolve in a more complex way.
To describe such situations,
we denote,
by $n(u)$,
the typical strength of the bonds (or the number of bonds)
at the layer specified by the non-positive integer $u$.
For the MERA for CFTs, $n(u)$ does not depend on $u$.
The MERAs for non-CFTs
are expected to correspond to the gravity dual with the metric
\begin{align}
  ds^2=g_{uu}du^2+\f{e^{2u}}{\ep^2}d\vec{x}^2+g_{tt}dt^2.
  \label{gmetr}
\end{align}
The metric for $\vec{x}$ is fixed because we employ
the same coarse-graining procedure.
The metric $g_{uu}$ in the extra dimension is related to the density of disentanglers,
i.e., $n(u)$ as we explain soon later.

The temporal component of the metric $g_{tt}$ cannot be determined from the
information of entanglement entropy $S_A$.
If we generalize the definition of the entanglement entropy so that
the subsystem $A$ is defined on a Lorentz boosted time slice,
in principle, we can read off $g_{tt}$,
although we will not discuss this issue in detail in this paper.

\begin{figure}[ttt]
   \begin{center}
     \includegraphics[height=4cm]{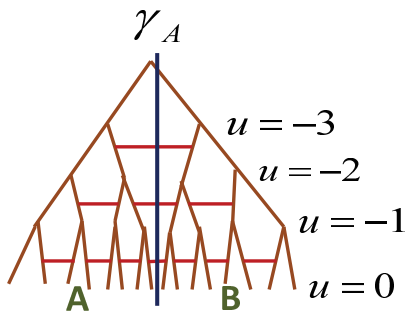}
   \end{center}
   \caption{
     The calculation of the entanglement entropy $S_A$ in the
     MERA for a one dimensional system,
     when the subsystem $A$ is the half of the total space.
   }\label{fig:ABE}
\end{figure}

To relate the MERA to the geometry (\ref{gmetr}),
let us calculate the entanglement entropy $S_A$ when $A$
is the half of the space
(i.e., we consider the equal bipartition of the total space).
The calculation in the MERA
when the system is a one-dimensional spin chain
is depicted in Fig.\ \ref{fig:ABE}.
In the MERA for a $(d+1)$-dimensional quantum system,
following the previous discussion,
we can estimate the entanglement entropy as
\begin{align}
  S_A\propto L^{d-1}\sum_{u=-\infty}^0 n(u)\cdot 2^{(d-1)u},
  \label{sanu}
\end{align}
where $L^{d-1}$ is the number of lattice points on the boundary of $A$.
On the other hand,
the holographic formula in AdS/CFT (\ref{HEE}) leads to
\begin{align}
  S_A=\f{1}{4G_N}\cdot \f{V_{d-1}}{\ep^{d-1}}\int^{u_{\mathrm{UV}}(=0)}_{u_{\mathrm{IR}}(=-\infty)}du \s{g_{uu}}e^{(d-1)u},
  \label{sahol}
\end{align}
where $V_{d-1}$ is the area of the boundary of $A$.
In the continuum limit $\ep\to 0$, using the obvious relation
$V_{d-1}=L^{d-1}\ep^{d-1}$, we find that (\ref{sanu}) and (\ref{sahol})
have the similar structure;
The comparison between the two calculations suggest
that
the layer index $u$ in the MERA and
the radial coordinate $u$ AdS/CFT are related as
$u_{\mathrm{AdS/CFT}}=\log 2\cdot u_{\mathrm{MERA}}$,
and moreover,
\begin{align}
  n(u)\propto \f{\s{g_{uu}}}{G_N}
  \label{disden}.
\end{align}
This implies that we can interpret the MERA
in the general setups as the holography.

The arguments so far are based on the MERA for discrete quantum systems. In order to make the connection to {AdS/CFT} more precise, it would be desirable to take the continuum limit from the
beginning by employing the idea of the cMERA \cite{cMERA} and this will be the main purpose in coming subsections. Especially we would like to give an interpretation of the cMERA in terms of holography by defining an appropriate metric $g_{uu}$ from the cMERA for any given quantum field theory. Before we go on to the construction of metric in the cMERA,
we would like to introduce
a natural metric in quantum mechanics.

  We close this subsection with one important comment.
In AdS/CFT, we need to take the large $N$ and strong coupling limit of gauge theories in order to realize the classical gravity limit (or equally Einstein gravity limit) where the holographic formula (\ref{HEE}) can be applied. If we abandon the strong coupling limit, we expect higher derivative corrections to the Einstein gravity and the holographic entanglement entropy
also includes higher derivative terms \cite{MyH,BKP}. If we do not take the large $N$ limit, the
gravity theory receives substantial quantum gravity corrections and the effective gravity action
will become highly non-local, for which the holographic entanglement entropy has not been calculated at present. Therefore, one may wonder how these two limits can be seen in
the MERA. Though we are not going to address a definite answer to this question, we can suggest a related important idea. In order to justify the identification (\ref{bondar}), we need to assume that the all relevant bonds are (almost) maximally entangled.
If this is not the case,
the precise estimation of the entanglement entropy gets
quite complicated,
and we need the information of entanglement of bonds which is far from the minimal
surface $\gamma_A$.
Therefore, in such situations,
calculations are expected to be ``non-local''
in the sense of tensor network geometry,
which is identified with a AdS space.
This may correspond to the
fact that the bulk gravity becomes non-local
if we do not take the large $N$ limit.
At the same time, it is natural to expect that the description
in terms of maximally entangled bonds can be valid in this limit.
We would like to leave more understandings for an important future problem\footnote{
After the preprint of this paper appeared on the arXiv, we noticed the preprint \cite{SWN}, where interesting behaviors in the large N limit of MERA were discussed.}.

\subsection{Quantum Distance}

Suppose we are given two quantum states in the Hilbert space.
They can be pure or mixed states.
For our discussion below, we only need pure states.
However, let us be general for a moment, and consider two mixed states,
characterized by a density matrix $\rho_{1}$ and $\rho_{2}$, respectively.
To measure how different these states are,
we can introduce a proper distance.
The Bures distance is given by
\begin{align}
D_{\mathrm{B}}(\rho_1,\rho_2)
&:=
2 \left(
  1 - \mathrm{Tr}\, \sqrt{
    \rho^{1/2}_1 \rho^{\ }_2 \rho^{1/2}_1
  }
\right).
\end{align}
Alternatively, one can consider the Hilbert-Schmidt distance
defined by
\begin{align}
D_{\mathrm{HS}}(\rho_1,\rho_2)
&:=
\frac{1}{2}
\mathrm{Tr}\, \left(
  \rho_1 -\rho_2
\right)^2.
\end{align}
When $\rho_1$ and $\rho_2$ are pure,
and given by
$\rho_1=|\psi_1\rangle \langle \psi_1|$
and
$\rho_2=|\psi_2\rangle \langle \psi_2|$,
respectively,
these definitions reduce to
\begin{align}
  D_{\mathrm{B}}(\rho_1,\rho_2)
  &=
  D_{\mathrm{B}}(\psi_1,\psi_2)
  =
  2\left(
    1 - | \langle \psi_1|\psi_2 \rangle|
  \right),
  \nonumber \\
  D_{\mathrm{HS}}(\rho_1,\rho_2)
  &=
  D_{\mathrm{HS}}(\psi_1,\psi_2)
  =
  1 - | \langle \psi_1|\psi_2 \rangle|^2.
  \label{DB, DHS for pure}
  \end{align}

Similarly,
for a set of quantum states
$\{\rho(\xi) \}$
parameterized by
finite numbers of real parameters $\xi=(\xi_1,\xi_2,\ldots)$,
we can define the distance between them
by
$D_{\mathrm{B}}[\rho(\xi), \rho(\xi+d\xi)]$
or
$D_{\mathrm{HS}}[\rho(\xi), \rho(\xi+d\xi)]$.
Assuming
the quantum state $\rho(\xi)$ changes smoothly as a function of
$\xi$,
we can expand
$D_{\mathrm{B}/\mathrm{HS}}[\rho(\xi), \rho(\xi+d\xi)]$
for infinitesimal $d\xi$.
Specializing to the case of pure states,
$\rho(\xi)=|\psi(\xi)\rangle \langle \psi(\xi)|$,
\begin{align}
  D_{\mathrm{B}}[ \psi(\xi),\psi(\xi+d\xi)  ]
  &= D_{\mathrm{HS}}[ \psi(\xi),\psi(\xi+d\xi)  ]
  =g_{ij}(\xi)d\xi_i d\xi_j,
  \label{metq}
\end{align}
where we have introduced the quantum metric tensor (or equally Fisher information metric) by
\begin{align}
g_{ij}(\xi) =
\mathrm{Re}\,
\langle \partial_i \psi(\xi) | \partial_j \psi(\xi) \rangle
-
\langle \partial_i \psi(\xi) |\psi\rangle \langle \psi| \partial_j \psi(\xi) \rangle.
\end{align}
$g_{ij}(\xi)$ transforms in a covariant fashion under reparameterization,
$\xi\to \xi'(\xi)$,
$g_{ij}(\xi)$
$=$
$(\partial \xi'_k/\partial \xi_i)
(\partial \xi'_l/\partial \xi_j)$
$
g_{kl}(\xi').
$

As an example, let us consider
a system with two bosonic oscillators defined by
creation and annihilation operators $(a,b)$ and $(a^\dagger, b^{\dagger})$.
They satisfy the commutation relations
$[a,a^\dagger]=[b,b^{\dagger}]=1$.
(By identifying
$a=a_k$ and $b=a_{-k}$,
we can directly apply calculations here
to those for the free scalar field for a fixed $k$ sector.)
Then we can explicitly write the state defined  (\ref{conda}) as
\begin{align}
|\psi_{\lambda}\lb=\s{1-|\lambda|^2}\cdot e^{-\lambda a^+b^{+}}
|0\rangle,
\end{align}
where $\lambda$
is related $A_k$ and $B_k$ via $\lambda=B_k/A_k$.
Then the
inner product is found to be \be \la
\psi_{\lambda'}|\psi_{\lambda}\lb
=\f{\s{(1-|\lambda|^2)(1-|\lambda'|^2)}}{1-\lambda'^*\lambda}. \ee
By taking the limit
$d\lambda=\lambda'-\lambda \to 0$,
the quantum metric is computed as
\begin{align}
ds^2_k=\f{d\lambda d\lambda^*}{\left(1-|\lambda|^2\right)^2}=(A_k dB_k-B_kdA_k)(A^*_k dB^*_k-B^*_kdA^*_k), \label{infm}
\end{align}
where we used the relation $|A_k|^2-|B_k|^2=1$. This metric (\ref{infm}) coincides with that of the two dimensional hyperbolic space or Euclidean AdS$_2$.

\subsection{Emergent Metric from cMERA}

Now we consider the construction of the metric in the cMERA.
Again we assume the translational symmetry in the $x$ direction and would
like to construct $g_{uu}$ from the data in the cMERA.
We argue that a natural
definition of the metric component $g_{uu}$ for static states in the cMERA is obtained by applying the quantum metric (\ref{metq}) in an
appropriate way as described below. Since the Hilbert space of scalar field theory consists of the direct product of each sectors with fixed momentum $k$, finally we need to integrate over $k$. We define the metric $g_{uu}$ in the cMERA as follows:
\be
g_{uu}(u)du^2
={\cal N}^{-1}\left(
  1-\left|\la \Psi(u)|e^{iL\cdot du}|\Psi(u+du)\lb\right|^2\right)
={\cal N}^{-1}\left(1-\left|\la \Phi(u)|\Phi(u+du)\lb\right|^2\right), \label{rmn}
\ee
where we employed the relation (\ref{vy}). The normalization factor ${\cal N}$ is given by the
volume of the effective phase space at length scale $u$:
\be
{\cal N}=\mbox{Vol}\cdot \int_{|k|\leq \Lambda e^{u}}dk^d, \label{norma}
\ee
where Vol is the (infinite) volume of $d$ dimensional space $R^d$.

Notice that this measures the density of strength of the unitary transformations induced by disentanglers.
To extract
only the effect of disentanglers we need to get rid of the scale transformation (or coarse-graining) by putting the operator $e^{iL\cdot du}$ in (\ref{rmn}).
Comparing with (\ref{DB, DHS for pure}),
we have normalized the metric by
the volume of the phase space at the scale $u$
in the denominator so that the metric measures
the density of the strength of disentanglers. Note that it is still possible to multiply any finite constant with the definition (\ref{rmn}). This normalization turns out to be convenient when comparing the quantum metric in the MERA with the bulk background metric in AdS/CFT.
Observe that (\ref{rmn}) can be rewritten as\footnote{It would also be possible to introduce
the metric as
$
g_{uu}(u)du^2
={\cal N}^{-1}\left(
1-|\langle \tilde{\Psi}(u)|\tilde{\Psi}(u+du) \rangle|^2\right)
$
with
$
|\tilde{\Psi}(u) \rangle
:=
U(0,u)|\Omega\rangle.
$
This definition of the metric leads to
qualitatively similar results.
}
\ba
 g_{uu}(u)=\la
\Phi(u)|\hat{K}(u)^2|\Phi(u)\lb-\la
\Phi(u)|\hat{K}(u)|\Phi(u)\lb^2.
\ea

The reasoning to identify the quantum metric
(\ref{rmn}) as
a bulk (or holographic)
metric is based
on their relation to the entanglement entropy,
as we discussed for the discrete MERA
(see around (\ref{sanu})-(\ref{sahol}));
when we choose the subsystem $A$ to be the half of the total space,
the disentanglers
``cut'' by the division $\gamma_A$ of the MERA
network, which is induced by the bipartitioning in
defining the entanglement entropy,
contribute to the entanglement entropy.
The strength of the disentangler is a function of $u$.
The quantum metric  $\s{g_{uu}}$ measures
the density of the strength of disentanglers.
On the other hand,
in the classical gravity limit of AdS/CFT,
the holographic formula (\ref{HEE})
relates the area of the minimal surface ($\gamma_A$)
and the entanglement entropy.
These considerations support the identification (\ref{rmn}) as
the bulk metric.

In time-dependent states such as the ones obtained from the quantum quenches, we need to
be careful to apply the formula (\ref{rmn}) due to the ambiguity of the phase factor $\theta_k$ defined in (\ref{abqq}) noted in previous section. More details of this will be discussed in the next subsection.\footnote{We present another definition of metric in appendix \ref{another}. This is free from this ambiguity even though it can only be applicable to coherent states. This metric coincides with (\ref{rmn}) if the function $g(u)$ is real.}

Also we need to remember that the use of (\ref{HEE})
is restricted to the case where we take the strongly coupling and large $N$ limit of the CFT. In the free scalar field theory model where we perform detailed calculations, the dual gravity is expected to be highly quantum and the metric $\s{g_{uu}}$ cannot be regarded as that of the classical Einstein gravity. Nevertheless, we can still propose that (\ref{rmn}) always provides a metric which describes the dual geometry in a qualitative way.
Indeed, as the calculations of the entanglement entropy in
the 4D $\mathcal{N}=4$ super Yang-Mills theory
suggest
\cite{RT},
in some cases, the behavior of the entanglement entropy
does not qualitatively change
when we dial the coupling constant from zero to infinity.
It would be an very interesting future problem to somehow
explore the metric $\s{g_{uu}}$ in strongly coupled large $N$ gauge theories
to see a direct connection to the Einstein gravity.

\subsection{Coordinate Transformations and Time-Dependent Backgrounds}

A very important question on the interpretation of AdS/CFT in terms of the cMERA is how the
diffeomorphism invariance in the gravity is encoded in the cMERA. Even though we
leave comprehensive understandings of this issue for future works, here we would like to
try to present a schematic explanation. As we assumed in the formula (\ref{rmn}), we impose the translation symmetry in $\vec{x}$ direction for the states we consider.

First of all, we notice that the change of the cut off function
$\Gamma(|k|e^{-u}/\Lambda)$ can be understood as a part of the diffeomorphism. This is because if we change $\Gamma(|k|e^{-u}/\Lambda)$ into
$\Gamma(|k|\eta(e^{-u})/\Lambda)$, then this corresponds to the coordinate transformation $e^{-u}\to \eta(e^{-u})$ in the
extra dimension.

Next, remember that there is an ambiguity of choices of the phase factor $\theta_k$ (\ref{abqq}) in excited states as we noted in the previous section. Though the UV state $|\Psi\lb$ is clearly independent of $\theta_k$, the states $|\Psi(u)\lb$ at intermediate energy scale depend
on $\theta_k$ in general.
Note that $\theta_k$ can be an arbitrary function of time $t$.
It is tempting to argue
that this ambiguity corresponds to the choices of time slices in the bulk gravity. The time slice $t=t_0$ in a quantum field theory just corresponds to the one at the boundary of the gravity dual
(e.g., $z=0$ in the AdS metric (\ref{adsmet})),
following the standard idea of holography.
In the gravity dual, we need to extend the time slice at the boundary into the bulk in order to
find the gravity dual of a series of states $|\Psi(u)\lb$. It is clear that there are infinitely different ways to do this extension. In a translationally invariant background, this bulk time slice at time $t_0$ is specified by $F(t,z)=t_0$ in terms of the bulk time $t$ and radial coordinate $z(=\ep e^{-u})$, for a certain function $F$. The phase ambiguity $\theta_k(t)$ indeed has the same degree of freedom as $F(t,z)$ by roughly identifying $z=1/k$. It is curious to notice that the ambiguity $\theta_k$ originally comes from the fundamental fact that the state or wave function in quantum mechanics is only defined up to a phase factor.

Now we would like to consider how we can apply the metric (\ref{rmn}) to time-dependent states in the cMERA. For this, it is useful to remember the holographic calculations of
the entanglement entropy in time-dependent backgrounds \cite{HRT}, where $S_A$ is given by
the area of extremal surface in the full Lorentzian spacetime instead of a minimal surface on a time-slice. By considering the case where $A$ is a half space and the holographic calculation (\ref{sahol}) as before, we can argue that the metric (\ref{rmn}) leads to the correct metric in the gravity dual if we choose the phase $\theta_k(t)$ such that the extremal surface $\gamma_A$ is on the corresponding time slice.

As a consistency check of the above argument, we can explicitly confirm that only for the ground states in free scalar field theories, the metric component (\ref{rmn}) does not depend on the choice of $\theta_k$. In this static case, the metric corresponds to the the most natural time slice $t=t_0$ in the coordinate (\ref{gmetr}).

\subsection{Calculation of Metric in Free Scalar Field Theory}

To calculate the metric $g_{uu}(u)$ explicitly,
we focus on the free scalar field theory with a mass $m$.
Let us consider a class of states which are given by the coherent states
\be
(A_k(u)a_k+B_k(u)a^{\dagger}_{-k})|\Phi(u)\lb=0,
\ee
where $A_k(u)$ and $B_k(u)$ are related to
$M_k(u)$ via (\ref{xc}).
By using (\ref{infm}), (\ref{xc}), (\ref{gks})
and (\ref{uuu}),
the metric is computed as
\begin{align}
  g_{uu}(u)&=
  {\cal N}^{-1}\int_{|k|\leq \Lambda e^u} d^d k\,
  \left[ A_k(u)\de_uB_k(u)-B_k(u)\de_uA_k(u)\right]
  \left[ A^*_k(u)\de_uB^*_k(u)-B^*_k(u)\de_uA^*_k(u)\right]
\nonumber \\
&=
{\cal N}^{-1}\int_{|k|\leq \Lambda e^u} d^d k\,
\left[A_k(u)^2g(u)-B_k(u)^2g^*(u)\right]
\left[A^*_k(u)^2g^*(u)-B^*_k(u)^2g(u)\right],
\label{guuab}
\end{align}
where ${\cal N}$ was the normalization factor defined in (\ref{norma}).

Let us focus on the special case where $g(u)$ defined in (\ref{gks}) is real. Then
(\ref{msde}) and (\ref{xc}) tell us that $M_k(u)$, $A_k(u)$ and $B_k(u)$ are also real valued.
Under this assumption,
the expression (\ref{guuab}) is simplified as
\be
g_{uu}(u)= g(u)^2.  \label{fff}
\ee

\subsubsection{Ground states}

If we consider the ground state in the free scalar field theory, we find (\ref{fff}) leads to
\be
g_{uu}(u)=\chi(u)^2=\f{e^{4u}}{4(e^{2u}+m^2/\Lambda^2)^2}.
\ee
We introduce a new coordinate $z$ by
\be
e^{2u}=\f{1}{\Lambda^2z^2}-\f{m^2}{\Lambda^2},
\ee
where $0<z<1/m$.
Then the metric looks like
\be
ds^2=\f{dz^2}{4z^2}+\left(\f{1}{\Lambda^2z^2}-\f{m^2}{\Lambda^2}\right)dx^2+g_{tt}dt^2.  \label{methol}
\ee
This metric is capped off (vanishes)
at $z=1/m$ and this is consistent with the mass gap
in the scalar field theory.

In the massless case $m=0$, the spacial part of the metric (\ref{methol}) coincides with that of the pure AdS.
Also for the Lifshitz-like critical theory, we can show that the metric $g_{uu}$ is a constant as follows
from (\ref{mlif}). This is consistent with the proposed metric dual to Lifshitz-like fixed points \cite{Lif}.

\subsubsection{Quantum Quenches}

Now we would like to consider the time-dependent metric which corresponds
to the excited state
after a global quantum quench. The calculations get highly complicated as we need to specify the phase $\theta_k$ so that the
time-slice coincides with the extremal surface
$\gamma_A$ as we explained. Here we would like to do a shortcut by fixing $\theta_k$
artificially. This is still enough to obtain qualitative behavior of the time-dependence
of the metric. The metric can be obtained from the formula
(\ref{fff}) by substituting (\ref{mqc}) and (\ref{abq}).

First of all, it is clear from (\ref{mqc}),
the square root of the metric $\s{g_{uu}}=g(u)$ in (\ref{rmn})
grows linearly in time $t$.
This agrees with the fact that the entanglement entropy increases
linearly in time $t$ after quantum quenches,
which has been shown first for two dimensional CFTs in \cite{Quench},
and has been later obtained holographically in any dimensions in
\cite{Quenchhol}.
This offers a non-trivial evidence for
our proposed metric (\ref{rmn}).

Next let us look at Fig.\ \ref{fig:QunechMERA} again. This is now interpreted as the plot of $\s{g_{uu}}$ as a function of the radial coordinate $z(=1/|k|)$ and $t$ in the gravity dual. We can observe that the quantum quench induces the gravitational waves approximately within the light-cone $z<t$, which agrees with the causality in the bulk.

\subsubsection{Towards Flat Space Holography}

Let us assume the behavior \be g_{uu}(u)=g(u)^2\propto e^{2w u}, \label{hmets} \ee
where $w$ is a constant. For a generic form of the dispersion relation
$\ep=\ep_k$, we find the relation
(\ref{gdp}) between $g(u)$ and $\ep_k$. Therefore the metric
(\ref{hmets}) corresponds to the dispersion relation \be \ep_k\propto e^{A\cdot
k^{w}}, \ee where $A$ is a certain positive constant. This
dispersion relation is unusual and is highly non-local. Indeed, the
corresponding Hamiltonian looks like
\be
H=\int d^d x\,
\phi(x)e^{A(-\de^2)^{w/2}}\phi(x).
\ee

The metric (\ref{gmetr}) with (\ref{hmets}) becomes almost flat on a time slice
if we choose  $w=1$. This corresponds to the Hamiltonian
\be
H=\int d^d x\,
\phi(x)e^{A\s{-\de^2}}\phi(x).
\ee
Notice that a very similar non-local
field theory appears in \cite{WT} where a possible dual to gravity in
flat space has been explored.
A lesson we obtained here is that when we consider the holographic dual of
the metric (\ref{hmets}) with $w$ non-vanishing, the dual theory is expected to be
highly non-local.

\section{Another example: Free Fermion in Two Dimensions}

So far we have considered the free scalar field theory as an example.
Here we would like to briefly show that similar calculations can be done for free fermions.
 The cMERA for the free fermion has been worked out in the version 1 of
\cite{cMERA} and we follow almost the same convention. Readers who are interested in general discussions may skip this section and move on to the final
section.

For simplicity,
we focus on the free Dirac fermion $\psi$ in $(1+1)$ dimensions,
defined by
the action
\be
S_F=\int dtdx \left[i\bar{\psi}(\gamma^t\de_t+\gamma^x\de_x)\psi-m\bar{\psi}\psi\right],
\ee
where $\psi=(\psi_1,\psi_2)^T$ is
the two-component complex fermion
and $\gamma$ matrices are defined by
$\gamma^t=\sigma_3$ and $\gamma^x=i\sigma_2$
in terms of the Pauli matrices. Also we define
$\bar{\psi}=\psi^{\dagger}\gamma^t$ as usual.
The Hamiltonian of this theory is given by
\begin{align}
  H&=\int dx \left[-i\bar{\psi}\gamma^x\de_x \psi+m\bar{\psi}\psi\right]
  \nonumber \\
&=\int dk\left[k \psi _{1}^{\dagger} (k)\psi_{2} (k)+k \psi
_{2}^{\dagger} (k)\psi_{1} (k)+m \psi _{1}^{\dagger} (k)\psi_{1} (k)
-m\psi _{2}^{\dagger} (k)\psi_{2} (k) \right ],
\end{align}
where we performed the Fourier transformation. The canonical quantization leads to the following anti-commutation relation
\be
\{\psi_1(k),\psi^{\dagger}_1(k')\}=\{\psi_2(k),\psi^{\dagger}_2(k')\}=\delta(k-k').
\ee

\subsection{cMERA for Free Fermion}

The unentangled IR state can be defined by
\begin{equation}
\psi_{1}(k) \left| \Omega\right \rangle =0, \quad
\psi_{2}^{\dagger}(k) \left| \Omega\right \rangle =0.
  \label{IR}
\end{equation}
On the other hand, the true ground state $|\Psi\lb$ of this free
fermion theory is given by
\begin{equation}
\tilde{\psi}_{1}(k) \left| \Psi\right \rangle =0,
\quad
\tilde{\psi}_{2}^{\dagger}(k) \left| \Psi\right \rangle =0,
\end{equation}
where $\tilde{\psi}_{1,2}$ is defined by
\begin{equation}
\begin{split}
\tilde{\psi}_{1}(k)&=A^{(0)}_k\psi_1 (k)+ B^{(0)}_k\psi_2(k) \\
\tilde{\psi}_{2}(k)&=-B^{(0)}_k\psi_1 (k)+ A^{(0)}_k\psi_2(k),\\
\end{split}
\end{equation}
with
\be
A_k^{(0)}=\f{-k}{\s{k^2+(\s{k^2+m^2}-m})^2},\ \ \ \ B_k^{(0)}=\f{m-\s{m^2+k^2}}{\s{k^2+(\s{k^2+m^2}-m})^2}. \label{abcof}
\ee
These fermion operators satisfy the following commutation relations:
\begin{equation}
\begin{split}
\left[ H, \tilde{\psi}^{\dagger}_1(k)\right]=&\sqrt{k^2+m^2}\tilde{\psi}^{\dagger}_1(k),\\
\left[ H, \tilde{\psi}^{\dagger}_2(k)\right]=&-\sqrt{k^2+m^2}\tilde{\psi}^{\dagger}_2(k).\\
\end{split}
\end{equation}
More generally we can consider the excited states defined by
\begin{equation}\label{excite}
\begin{split}
\left( A_k\psi _1(k)+B_k \psi _2(k)\right)\left| \Psi \right \rangle &=0,
\\
\left( -B_k\psi _1^{\dagger}(k)+A_k \psi _2^{\dagger}(k)\right) \left| \Psi \right \rangle &=0.\\
\end{split}
\end{equation}
We can normalize
\begin{equation}
\left| A_k \right|^2 +\left| B_k \right|^2 =1.
\end{equation}
Notice that the state $|\Psi\lb$ does not change if we multiply a common phase factor to
$(A_k,B_k)$ as in the previous scalar field example.
The choice $(A_k,B_k)=(A^{(0)}_k,B^{(0)}_k)$ corresponds to the ground state.

As before, we aim
to relate the UV state $\left| \Psi \right \rangle$ to the
common IR state via the unitary transformation (\ref{xxx}).
We assume the following form for the disentanglers
\begin{equation}
  \hat{K}(u)=i\int dk\left[ g_k\left( u \right)
    \psi_1^{\dagger}(k)\psi_2(k)+g^{*}_k\left( u \right)\psi_1(k)\psi_2^{\dagger}(k)\right],
\end{equation}
where the function $g_k(u)$ is chosen to be the form
\begin{equation}
g_k(u)=g(u)\Gamma( |k| e^{-u}/\Lambda) \f{ ke^{-u}}{\Lambda}.
\end{equation}
where $g(u)$ is complex-valued in general.
The presence of the factor $k$, which is
missing in the scalar field theory, is now required to reproduce the correct ground state (\ref{abcof}) as noted in \cite{cMERA}.

In terms of the particle
annihilation and anti-particle creation operators, we obtain
\begin{align}
  &
  P e^{ -i\int^{u}_{u_{\mathrm{IR}}}\hat{K}(u')du' }
 \begin{pmatrix}
   \psi_1(k) \\
   \psi_2(k)
 \end{pmatrix}
 \tilde{P} e^{ i\int^{u}_{u_{\mathrm{IR}}}\hat{K}(u')du'}
   =
   M_k(u)
 \begin{pmatrix}
 \psi_1( k) \\
 \psi_2( k) \\
\end{pmatrix},
\nonumber \\
&
 P e^{ -i\int^{u}_{u_{\mathrm{IR}}}\hat{K}(u')du'}
 \begin{pmatrix}
   \psi_1 ^{\dagger}(k)
   \\
 \psi_2 ^{\dagger}(k)\\
\end{pmatrix}
\tilde{P} e^{ i\int^{u}_{u_{\mathrm{IR}}}\hat{K}(u')du'}
=
N_k(u)
 \begin{pmatrix}
 \psi_1^{\dagger}( k) \\
 \psi_2^{\dagger}( k) \\
 \end{pmatrix},
\end{align}
where we introduced rank 2 matrices $M_k(u)$ and $N_k(u)$
as
\begin{align}
M_k(u)&\equiv
\begin{pmatrix}
P_k(u) &Q_k(u) \\
-Q^{*}_k(u) &P^{*}_k(u) \\
\end{pmatrix}
=\tilde{P}\exp{\left(\int^{u}_{u_{\mathrm{IR}}}du'G_k(u')\right)},
\nonumber \\
N_k(u)&\equiv
\begin{pmatrix}
P^{*}_k(u) &Q^{*}_k(u) \\
-Q_k(u) &P_k(u) \\
\end{pmatrix}
=\tilde{P}\exp{\left(\int^{u}_{u_{\mathrm{IR}}}du'H_k(u')\right)},
\end{align}
with $G(u)$ and $H(u)$ defined by
\begin{equation}
  \label{parameter}
  G_k(u)
    =
\begin{pmatrix}
0 & -g_k(u)\\
g^{*}_k(u) &0 \\
\end{pmatrix},
\quad
H_k(u)=
\begin{pmatrix}
0 &-g^{*}_k(u)\\
g_k(u) &0 \\
\end{pmatrix}.
\end{equation}
Equivalently,we can show that
\begin{equation}
\frac{dM_k(u)}{du} =M_{k}(u)G_k(u),
\quad \frac{dN_k(u)}{du} =N_{k}(u)H_k(u).
\end{equation}
The matrices
$M_k(u)$ and $N_k(u)$ clearly satisfy $\left|P_k(u)
\right|^2+\left|Q_k(u)\right|^2=1$ and preserve both the commutation
relation of $\psi_1(k)$ and $\psi_1^{\dagger}(k)$ and the one of
$\psi_2(k)$ and $\psi_2^{\dagger}(k)$.
By comparing the IR and UV states, we obtain the relation
\be
P_k(0)=A_k(0),\ \ \ Q_k(0)=B_k(0).
\ee

Notice that $M_k(u)$ and $N_k(u)$ satisfy
\begin{align}
M_k(u)^{\dagger}M_k(u)=\text{\bf{1}},\quad \det M_k(u) =1,
\nonumber \\
N_k(u)^{\dagger}N_k(u)=\text{\bf{1}},\quad \det N_k(u) =1,
\end{align}
which says these belong to $SU(2)$. Also the parametrization (\ref{parameter}) covers only two out of three
generators of $SU(2)$. Though the remaining one is found to be
\begin{equation}\label{ig}
e^{i\delta_k \sigma_3} \in SU(2),
\end{equation}
we can ignore this by using the phase ambiguity of (\ref{excite}) as in the free scalar field
theory case.

\subsection{Ground State}

If we assume that $g_k(u)$ takes only real values i.e. $g_k=(g_k)^{*}$, then we find from (\ref{parameter})
\begin{equation}
M_k(u)=N_k(u)=
\begin{pmatrix}
\cos{\left(\int^{u}_{u_{\mathrm{IR}}}du'g_k(u') \right)}
&-\sin{\left(\int^{u}_{u_{\mathrm{IR}}}du'g_k(u') \right)} \\
\sin{\left(\int^{u}_{u_{\mathrm{IR}}}du'g_k(u') \right)}
&\cos{\left(\int^{u}_{u_{\mathrm{IR}}}du'g_k(u') \right)}
\end{pmatrix}.
\end{equation}
By comparing between these at the UV point $u=0$, we find
\begin{equation}
\sin 2\vp_k =\f{-k}{\s{k^2+m^2}}, \label{fsin}
\end{equation}
where we defined
\be
\vp_k=\int^0_{\log \left| k \right|/\Lambda} du g(u)\f{ke^{-u}}{\Lambda}.
\label{frel}
\ee
The function $g(u)$ is found from (\ref{frel}) as follows
\be
g(u)=-\f{k\left| k\right|}{\Lambda}\f{\de}{\de k}\left[\f{\Lambda}{k}\vp_k\right]\Biggl|_{\left| k\right| =\Lambda e^u}=\f{1 }{2}\left[-\arcsin\f{\Lambda e^u}{\s{\Lambda^2 e^{2u}+m^2}}+\f{m\Lambda e^u}{m^2+\Lambda^2 e^{2u}}\right].
\ee
In particular, in the massless case $m=0$ we find that $g(u)$ becomes a constant
\be
g(u)=\vp_k= -\f{\pi}{4}.
\label{fml}
\ee

\subsection{Metric from cMERA}

Now we would like to move on to a holographic interpretation
of the cMERA in this free fermion theory.
We can apply the general formula (\ref{rmn})
for the metric in the extra dimension.
The state defined by (\ref{excite}) is written as the coherent state
\begin{equation}
\left| \psi_{\lambda}\right \rangle= \frac{\left[1 -\lambda
\psi_1^{\dagger}(k)\psi_2(k) \right]}{\sqrt{1+\left| \lambda
\right|^2}} \left| \Omega \right \rangle ,
\end{equation}
where $\lambda=\frac{B_k}{A_k}$. We can show
\begin{equation}
\left \langle \psi_{\lambda '}\big{|}\psi_{\lambda} \right \rangle =
\frac{1+\lambda
\lambda'^{*}}{\sqrt{(1+\left|\lambda\right|^2)(1+\left| \lambda'
\right| ^2)}}.
\end{equation}
By considering an infinitesimal Bures distance (or equally Hilbert-Schmidt distance),
we find the corresponding metric.
If we take the limit $\lambda' \rightarrow \lambda$,
\begin{equation}
\begin{split}
  D_{\mathrm{B}}\left(
    \psi_{\lambda'} ,\psi_{\lambda}
    \right)&=\lim_{\lambda'\to \lambda}2\left(1-|\la \psi_{\lambda'}|\psi_{\lambda}\lb|\right)=\frac{d\lambda^{*}~d\lambda}{(1+\left| \lambda \right| ^2)(1+\left|\lambda' \right| ^2)} \\
&=(A_k^{*}dB_k^{*}-B_k^{*}dA_k^{*})(A_k dB_k-B_k dA_k) \\
\end{split}
\end{equation}
Finally the holographic metric (\ref{rmn}) is found to be
\begin{equation}
\begin{split}
  g_{uu}(u)&=
  \frac{1}{\mathcal{N}}
  \int_{|k|\leq \Lambda e^u} dk
  \left[
    A_k(u)^{*}\partial_u B_k(u)^{*}-B_k(u)^{*}\partial_u A_k(u)^{*}
  \right]
  \left[ A_k(u) \partial _u B_k(u)-B_k(u) \partial_u A_k(u) \right] \\
&=\frac{1}{\mathcal{N}}\int_{|k|\leq \Lambda e^u} dk
\left(\f{k^2e^{-2u}}{\Lambda^2}\right)
\left[
  A_k^2(u)
  g(u)+B_k(u)^2g(u)^{*} \right] \cdot
\left[
  A_k^{*}(u)^{2}
  g(u)^{*}+B_k^{*}(u)^{2}g(u)
\right], \no
\end{split}
\end{equation} where $\mathcal{N}$ is defined in (\ref{norma}).
When $g(u)$ is a real valued function, we find
\begin{equation}
g_{uu}(u)=\f{g(u)^2}{3}.
\end{equation}
For the ground state (\ref{fml}) of massless fermion theory, we can find that $g_{uu}$ is a constant and the gravity dual becomes a AdS space as expected from the conformal symmetry.

\section{Conclusions and Discussions}

By making use of the holographic formula for the entanglement
entropy as a hint,
we have introduced a proper metric to the MERA tensor networks
and in particular to their continuum counter parts in the cMERA.
The metric captures how quantum degrees of freedom are entangled
with each other at different length scales.
We computed the metric explicitly
for the cMERA for free boson and free fermion quantum field theories.
We have checked that the computed metric has properties
expected from the holography;
that it coincides with the AdS metric when these field
theories are conformal field theories;
that it is capped off when there is a mass gap;
that its square root grows linearly in time after quantum quench,
etc.

Our formulation based on the cMERA can be applied to holography for wide classes of spacetimes which are far more general than the AdSs. For example, we found that if we consider holographic duals of generic spacetimes such as a flat spacetime, they should be highly non-local theories including infinitely many derivatives in their Hamiltonians. It might be an interesting future problem to understand the physics of black holes in a flat spacetime by using this formulation. For example, they have the intriguing properties such as the negative specific heat. Interestingly, it has been known that the negative specific heat can happen in highly non-local theories \cite{NS}.
Also it is another important future problem to extend our formulation to the finite temperature
theories dual to AdS black holes (refer to the recent paper \cite{MIH} for a construction of black hole states in MERA based on the thermofield dynamics).

\begin{figure}[t]
   \begin{center}
     \includegraphics[height=4cm]{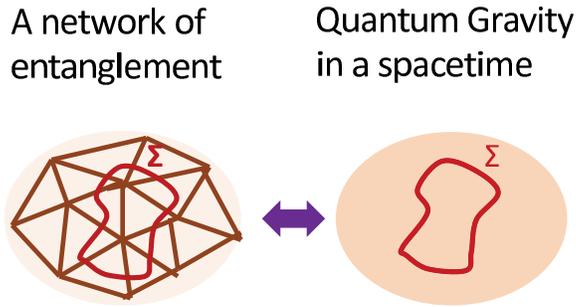}
   \end{center}
   \caption{A conceptual sketch of a possible formulation of quantum gravity in terms of entanglement.}\label{fig:CONJC}
\end{figure}

In the interpretation of the cMERA as AdS/CFT, the key identification is that each of entangling bonds (disentanglers) in an appropriate time slice can be regarded as a unit surface area. It is very exciting to note that this correspondence seems to be naturally generalized to spacetimes which have no space-like boundaries and which do not follow the standard rule of holography. This might offer us
a hint to formulate quantum gravity in general spacetimes (see Fig.\ \ref{fig:CONJC}). Notice that
in this correspondence, both sides are manifestly dynamical because the time coordinate is already built in and cannot be emergent. Refer to \cite{Ra,TNG} for closely related ideas which connect the quantum entanglement to the emergence of gravity dynamics.

A precise connection between the MERA and AdS/CFT is still missing, though.
Firstly, in our current approach, we could not determine the time component of
the holographic metric. In other words, our metric is the one on a fixed time slice
and this depends on the choice of the time slice. In principle we can boost the system and
calculate the entanglement entropy so that we can read off the time component of metric.
This certainly deserves future studies. Secondly, the role of the large $N$ limit in
the MERA is not clear yet
(although we gave a speculation on this point).
On the other hand,
the non-necessity of the large $N$ limit in formulating the MERA
can be thought of as an advantage;
our metric can, in principle, be defined for arbitrary quantum
many-body systems once their (ground) states are expressed in terms of
the MERA tensor networks.
Assuming that the MERA (and its suitable modification)
can be applied to a wide class of quantum many-body problems,
we can define a sort of bulk geometry averaged by quantum fluctuations for a wide class of systems.
It is also useful to note that the gravity dual of $O(N)$ scalar field theory is given by the
higher spin gravity \cite{Va} as has been recently studied actively \cite{KP,GY,KJJR,DMR,MaHS}. We will be able to employ this correspondence to study the relation between
the cMERA and AdS/CFT more closely. The excitations of higher spin fields will correspond to the addition of non s-wave terms to the disentanglers $\hat{K}$ in (\ref{dise}).

This mysterious connection between the MERA and AdS/CFT opens up a possibility to classify
various phases and states in quantum many-body systems
in terms of their bulk geometry.
This may be a good news,
since there is growing evidence that
quantum phases and quantum phase transitions
cannot fully classified in terms of the conventional tools
in classical statistical physics, such as the
Landau-Ginzburg type theories of phase transitions \cite{Wen}.
In the latter, classical phases can be characterized by
a pattern of symmetry breaking, i.e., in terms of group
theory.
On the other hands, there are known examples
that do not fall into this category;
Most notably, topological phases of matter such as the
fractional quantum Hall effect.
AdS/CFT and the MERA offer an emerging
holographic view for quantum phases,
which is complementary to
the Landau-Ginzburg paradigm,
viz,
quantum phases are to be classified by their emergent geometry.

\begin{figure}[t]
   \begin{center}
     \includegraphics[height=5cm]{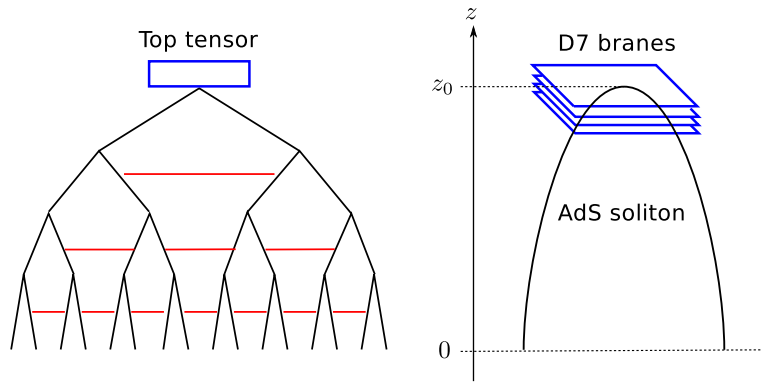}
   \end{center}
   \caption{
     A comparison between the MERA for topological phases
     and the gravity dual of the Chern-Simons theory.
\label{fig:toptensor}
}
\end{figure}

In order to offer further evidence for
the connection between {AdS/CFT} and the MERA
(``AdS/MERA'' in short),
it is worth while mentioning and comparing
how topological phases are represented
in the MERA and AdS/CFT.
(Fig.\ \ref{fig:toptensor}.)
In \cite{AguadoVidal08},
the MERA for a topologically ordered phase in two spatial
dimensions is constructed for a particular lattice model,
the Kitaev's toric code model.
One of the salient feature of the MERA tensor network for
the toric code model is the existence of its ``top tensor'',
which is located at the most IR region (``top'') of the
tensor network. The top tensor stores information
on the topological degrees of freedom (or topological order).
In particular, it captures one of the defining properties of
the topological phases, the topological ground state degeneracy
when the system is put on the Riemann surface of genus $g\ge 1$
(the ground state degeneracy is four in the toric code model
on the torus).
The tensor networks for the four ground states of the toric code
are largely identical except in their top tensor.

On the other hand,
in \cite{Fujita2009},
the gravity dual of a three dimensional gauge theory which flows in the IR to
the pure $SU(N)$ Chern-Simons theory is constructed.
The geometry is capped off at IR since the theory is gapped.
In addition, D7-branes are located at IR.
This setup is gapped even without the D7-branes, but
their presence is necessary to realize a topological phase;
the D7-branes give rise to the Chern-Simons term,
and contributes to the topological entanglement entropy.
The holographic views of topological phases offered from the MERA and
AdS/CFT are surprisingly similar, and
it is tempting to identity the top tensor and the D7-branes.

\section*{Acknowledgements}

We would like to thank Y.-C. Hu,
S. Flammia, Y. Hikida, R. Hubener, H. Matsueda, N. Ogawa, H. Ooguri,
X.-L. Qi, Y. Sekino, N. Toumbas and T. Ugajin for useful discussions. We are very grateful to the Aspen workshop
``Quantum Information in Quantum Gravity and Condensed Matter Physics,''
and the program ``Holographic Duality and Condensed Matter Physics'' at KITP in Santa Barbara where some parts of this work has been conducted. TT would like to thank the workshop ``Entanglement in quantum many body systems and renormalization,'' held in Yukawa Institute, Kyoto University for stimulating discussions on the entanglement renormalization. TT is also very grateful to the organizers and participants of the following meetings for illuminating discussions, where he gave presentations related to this paper: the theory workshop 2012 at KEK in Tsukuba, the conference ``Black Holes and Information,'' at KITP, the workshop ``discussion meeting on string theory'' at ICTS in Bangalore, the workshop ``Physics of information, information in physics, and the demon,'' at Institute for Molecular Science in Okazaki and the workshop ``Gravity Theories and their Avatars,'' held at Crete Center of Theoretical Physics in
 Heraklion. S.R. has been supported by the U.S. Department of Energy,
Office of Basic Energy Sciences, Division of Material Sciences
and Engineering (DE-FG02-12ER46875). TT is supported by JSPS Grant-in-Aid for Challenging
Exploratory Research No.24654057. TT is also
partially supported by World Premier International
Research Center Initiative (WPI Initiative) from the Japan Ministry
of Education, Culture, Sports, Science and Technology (MEXT).

\appendix

\section{Another Definition of Metric for Coherent States}\label{another}

Here we discuss an intuitive measure of entanglement and based on this we would like to define another metric
for the coherent states in the free scalar field theory. We consider the relative distance between the IR vector $(\ap_k,\beta_k)$ defined in (\ref{unents}) and UV one $(A_k,B_k)$ defined in
(\ref{conda}).  By performing the $SU(1,1)$ transformation, this is equivalent to
the distance between $(1,0)$ and $(A_k\ap_k-B_k\beta_k,\ap_kB_k-\beta_kA_k)$. Notice that in the UV limit $k=M$, the distance is vanishing. If the distance is small for finite $k$, then the UV state $|\Psi\lb$ is closer to the unentangled IR state $|\Omega\lb$. Therefore this distance presents the measure of entanglement.

Now we parameterize
\ba
&& A_k\ap_k-B_k\beta_k=\cosh a_k\cdot e^{ib_k},\no
&& \ap_kB_k-\beta_kA_k=\sinh a_k\cdot e^{ic_k}.
\ea
The natural metric of this parameter space which preserves $SU(1,1)$ structure is defined by \be
ds^2=da^2-\cosh^2 a db^2+\sinh^2 a dc^2,
\ee
which coincides with the $AdS_3$ as is expected from the $SU(1,1)$ symmetry.
Moreover, we need to take a quotient by the identification $(b,c)\sim (b,c)+(\theta,\theta)$ for any $\theta$, which is the phase ambiguity of the definition (\ref{conda}). In the end, we obtain the following AdS$_2$ metric \ba
ds^2=da^2+\cosh^2 a\sinh^2 a (db-dc)^2, \label{adstw}
\ea
and indeed it coincides with the quantum distance (\ref{infm}).

The vector $(a_k,b_k,c_k)$ for $0<k<M$ makes a trajectory $\Gamma_\Psi$ in this AdS$_2$ space. For simplicity we
start with a scalar field theory in $1+1$ dimension. The entanglement entropy $S_A$ for a given subsystem $A$ should be a functional of $\Gamma_\Psi$. A simplest choice will be just the length of $\Gamma_\Psi$.

The length $L_k$ of this trajectory is defined by
\be
L_k=\int^k_0 dk\s{(\de_k a_k)^2+\cosh^2 a_k\sinh^2 a_k(\de_k b-\de_k c)^2}.
\ee
$L_k$ describes how the entanglement is added as we increase the energy scale $k$.
 Therefore we are lead to identify
 \be
 \s{g^{E}_{uu}}=k\f{dL_k}{dk}=k\s{(\de_k a_k)^2+\cosh^2 a_k\sinh^2 a_k(\de_k b-\de_k c)^2}, \label{tmet}
  \ee
 assuming the relation $k=\Lambda e^{u}$. This defined a natural metric component $g^E_{uu}$
 in the extra dimension $u$.

Below we study the property of this metric in the massless free scalar theory $m=0$.
 For the vacuum state $|\Psi\lb$ i.e. $(A_k,B_k)=(1,0)$, we simply
find $g^{E}_{uu}=\f{1}{4}$.  Moreover, if $g(u)$ defined in (\ref{gks}) is real, we can show after some algebras
\be
g^{E}_{uu}=g(u)^2=g_{uu}.
\ee
Thus it agrees with the emergent metric defined in (\ref{rmn}). However, in general excited states this equivalence is no longer true.

It is also interesting to note that for the ground state $|\Psi\lb$ i.e. $(A_k,B_k)=(1,0)$ we find
\ba
L_{tot}\equiv L_{k=M}=\f{1}{2}\int^{M}_{\ep_{\mathrm{IR}}}\f{dk}{k}=\f{1}{2}\log\left(\f{M}{\ep_{\mathrm{IR}}}\right), \label{toth}
\ea
where $\ep_{\mathrm{IR}}$ is the IR cut off. This is proportional to the entanglement entropy
$S_A$ in two dimensions (i.e. $d=1$) when $A$ is the half of the total space $R^d$.

In general, we can express such a functional of $\Gamma_\Psi$ which is invariant under $SU(1,1)$
symmetry as
\be
S_A=\int_{\Gamma_\Psi}dk f(k)\s{G_{AdS2}}, \label{enten}
\ee
for any function $f(k)$, where $G_{AdS2}$ is the metric (\ref{adstw}). If we assume that $A$ is given by a half space of $R^{d}$, then the calculation should not depend on the energy scale i.e. $k$. Thus we can set $f(k)$ is a constant. This indeed agrees with the observation (\ref{toth}). On the other hand, when $A$ is a strip with a finite width $l$, we will need to choose $f(k)$ is some function of the form $f(k)=F(kl)$. To extend this argument to higher dimensions $d$, we just need to multiply the degeneracy factor $e^{(d-1)u}$ in front of (\ref{enten}).

\begin{figure}[ttt]
   \begin{center}
     \includegraphics[height=4cm]{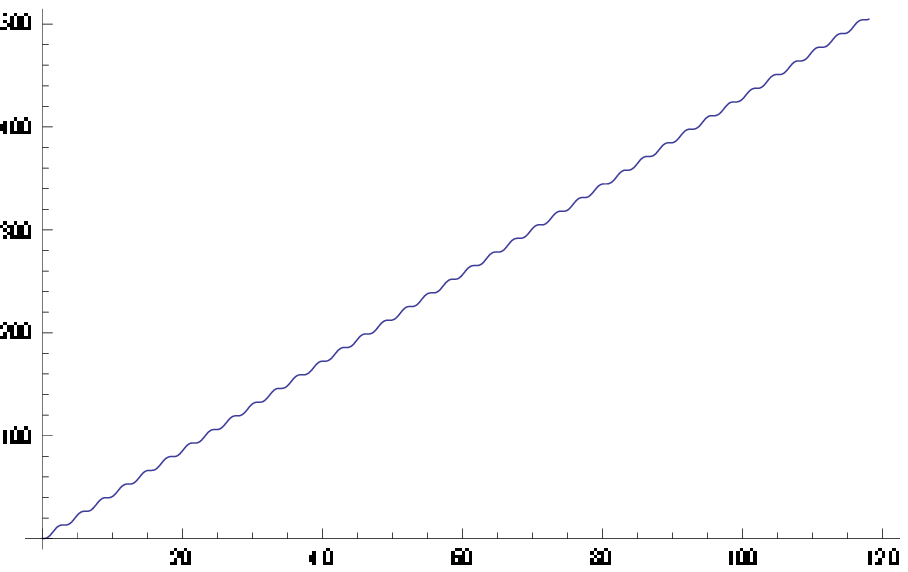}
     \hspace{1cm}
     \includegraphics[height=4cm]{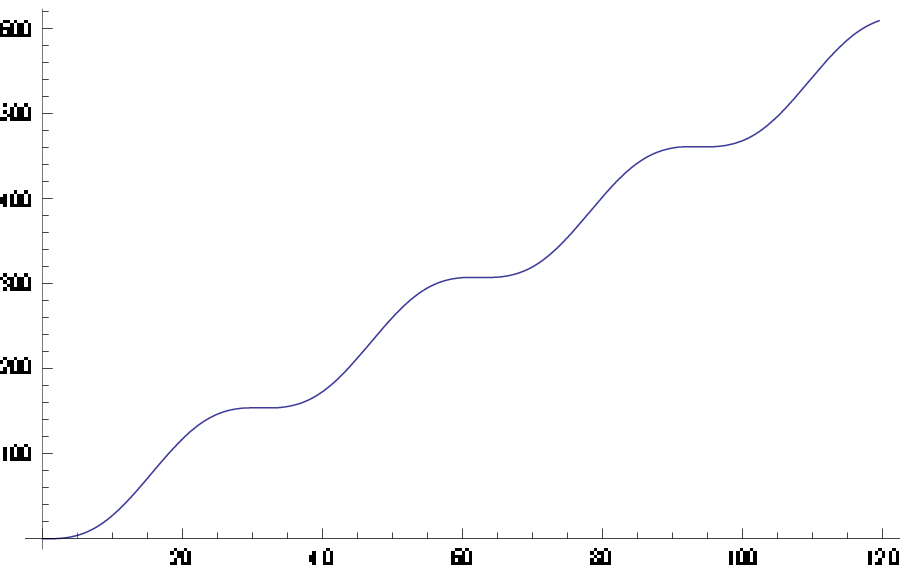}
   \end{center}
   \caption{Plots of $\s{g^{E}_{uu}}$ as a function of $t$. In the left graph we assume $z=1$, while the right one $z(=1/k)=10$. We assumed that $m_0=10$ and $m=0$ for this quantum quench.
    }\label{fig:quench2}
\end{figure}

Also we would like to analyze the global quantum quench where the mass of the free scalar changes
from $m_0$ to $m=0$. We plotted the time evolution of $\s{g^{E}_{uu}}$ for fixed values of $z$ in Fig.\ref{fig:quench2}. This behavior is clearly consistent with the linear growth of the entanglement entropy. We can also confirm that the plot $\s{g^{E}_{uu}}$ as a function $z=1/k$ and $t$ looks very similar to Fig.\ref{fig:QunechMERA}. We can again qualitatively confirm that the excitations are propagating within the light-cone $z<t$.

Finally notice that this definition of metric does not have any ambiguity due to the phase factor
as opposed to the one (\ref{rmn}) we proposed in the main text. However, $g^E_{uu}$ cannot be defined for general states as we employed the special property of coherent state that the state is written as the direct product of different momenta.

\end{document}